\documentclass[a4paper,11pt]{article}
\pdfoutput=1 

\usepackage{jcappub, tikz} 

\usepackage[T1]{fontenc} 
\usepackage[applemac]{inputenc}

\newcommand{\beq}{\begin{equation}}
\newcommand{\eeq}{\end{equation}}
\newcommand{\bfx}{{\boldsymbol{x}}}
\newcommand{\bfr}{{\boldsymbol{r}}}
\newcommand{\bfq}{{\boldsymbol{q}}}

\newcommand{\bftchi}{{\boldsymbol{\widetilde{\chi}}}}
\newcommand{\bfk}{{\boldsymbol{k}}}

\newcommand{\bfA}{{\boldsymbol{A}}}
\newcommand{\bfI}{{\boldsymbol{I}}}
\newcommand{\bfU}{{\boldsymbol{U}}}
\newcommand{\bfV}{{\boldsymbol{V}}}

\newcommand{\Mpc}{{\rm Mpc}}

\newcommand{\tdelta}{{\tilde \delta}}

\title{\boldmath On the primordial information available to galaxy redshift surveys}

\author{Matthew McQuinn}
\affiliation{Department of Astronomy, University of Washington,
 Seattle, WA 98195}
\emailAdd{mcquinn@uw.edu}

\abstract{
We investigate the amount of primordial information that can be reconstructed from spectroscopic galaxy surveys, as well as what sets the noise in reconstruction at low wavenumbers, by studying a simplified universe in which galaxies are the Zeldovich displaced Lagrangian peaks in the linear density field.  For some of this study, we further take an intuitive linearized limit in which reconstruction is a convex problem but where the solution is also a solution to the full nonlinear problem, a limit that bounds the effectiveness of reconstruction.  The linearized reconstruction results in similar cross correlation coefficients with the linear input field as our full nonlinear algorithm.  The linearized reconstruction also produces similar cross correlation coefficients to those of reconstruction attempts on cosmological N-body simulations, which suggests that existing reconstruction algorithms are extracting most of the accessible information.  Our approach helps explain why reconstruction algorithms accurately reproduce the initial conditions up to some characteristic wavenumber, at which point there is a quick transition to almost no correlation.  This transition is set by the number of constraints on reconstruction (the number of galaxies in the survey) and not by where shot noise surpasses the clustering signal, as is traditionally thought.  We further show that on linear scales a mode can be reconstructed with precision well below the shot noise expectation if the galaxy Lagrangian displacements can be sufficiently constrained.  We provide idealized examples of nonlinear reconstruction where shot noise can be outperformed.  
  }

\begin{document}
\maketitle
\flushbottom

\section{Introduction}

As we push within sight of mining the remaining cosmological information from the cosmic microwave background, there has been an increasing theoretical emphasis on understanding the late-time growth of cosmic structure as probed by galaxy redshift surveys \citep[e.g.][]{2007ApJ...664..660E, 2008MNRAS.389..497K, carrasco12, 2012JCAP...10..006T,2013MNRAS.432..894J, 2015MNRAS.446.4250A, mcquinn16, 2017MNRAS.469.1968P, 2017PhRvD..96b3505S, 2017JCAP...12..009S, modi18, 2018JCAP...07..043F, 2018arXiv180802002S}.   While we appear to have sufficient perturbative control to wavenumbers of $k\approx 0.15\;h\;$Mpc$^{-1}$ at $z\sim 0.5$ to obtain unbiased cosmological constraints (e.g., \cite{2017MNRAS.469.1968P, 2018arXiv181110640S, 2020arXiv200308277N}, although see \cite{2016arXiv160200674B}), it is unsettled whether galaxy surveys can extract useful constraints to higher wavenumbers.  Furthermore, at low wavenumbers it is unclear whether existing methods achieve the minimum possible error.  

The traditional approach to deriving cosmological bounds from large-scale structure measurements compares lower order statistics (such as the power spectrum and, perhaps, the bispectrum) measured from observational datasets with perturbation theory \citep[e.g.][]{bernardeau02, carrasco12, 2018arXiv181110640S}, large cosmological simulations \citep[e.g.][]{davis85, 2019arXiv190411923M, 2019MNRAS.485.3370G}, or quicker methods that take hybrid approaches \citep[e.g.][]{2013JCAP...06..036T, 2010ApJ...713.1322L, 2016MNRAS.463.2273F, 2019MNRAS.485.2407G}. One disadvantage of limiting the analysis to lower order statistics is that information becomes progressively entangled in higher order correlations with increasing wavenumber.  This entanglement has led to efforts that attempt to reconstruct the linear density field by in essence running the equations in reverse \citep{2007ApJ...664..660E, 2009PhRvD..79f3523P, 2012JCAP...10..006T}.  Such reconstruction methods have been applied to baryon acoustic oscillation (BAO) measurements in order to sharpen these features \citep{2014MNRAS.441.3524K, 2017MNRAS.464.3409B}, and they are anticipated to markedly improve the constraining power of future BAO surveys \citep{2007ApJ...664..675E}.

  Inspired by this success at reconstructing the BAO, a more recent focus has been on understanding the limits for how much primordial information can be extracted (rather than focus solely on sharpening the BAO).  One class of studies use perturbative bias expansions \citep{2017PhRvD..96b3505S, 2018arXiv180802002S, 2019JCAP...11..023M, 2019MNRAS.482.5685H, 2019JCAP...01..042S} and, an even more ambitious class attempts to compare fully nonlinear theories to data \citep{2008MNRAS.389..497K, 2013MNRAS.432..894J, 2015MNRAS.446.4250A,  2017JCAP...12..009S, 2018JCAP...07..043F, modi18}.  The simplest (albeit computationally intractable) formulation of the latter class would run simulations with all possible random initial condition fields, find the best match to the observations and, then, take the power spectrum of that simulation's density field to constrain the cosmology.  In practice, various optimizations have been devised such as assuming the displacement is a potential field and that the tracers were initially homogeneous \citep{2013ApJ...772...63W, 2017MNRAS.469.1968P}, improving MCMC algorithms so that the expensive posterior evaluations are more likely to be accepted \citep{2013ApJ...772...63W, 2015MNRAS.446.4250A}, using machine learning algorithms to extrapolate from a more-easily-simulated coarse density grid to something more akin to a halo field \citep{modi18}, evolving forward the density field with less expensive techniques than full simulations \citep{2015MNRAS.446.4250A, 2018JCAP...07..043F, 2019JCAP...11..023M} 
  and using methods that find some local posterior maximum and assuming/arguing that the biases from this being a local maximum are correctable \citep{2017JCAP...12..009S, 2018JCAP...07..043F}.  
    The efficacy of reconstruction tends to be similar regardless of methodology:  At low redshifts, reconstruction of the initial conditions is effective to wavenumbers of $k\approx 0.5~$Mpc$^{-1}$ given a dense enough galaxy survey, potentially extending the wavenumber reach over perturbative methods by a factor of a few \citep{2017MNRAS.469.1968P, 2013ApJ...772...63W, 2015MNRAS.446.4250A, 2018JCAP...07..043F}.  
  
  We will henceforth refer to these fully nonlinear efforts as ``reconstruction,'' even though the name was first coined in the context of spectroscopic galaxy surveys for reversing nonlinear evolution specifically in the BAO \citep{2007ApJ...664..675E}.  We further specialize to the class of reconstruction algorithms that start with a halo/galaxy field rather than the exploratory studies that reconstructed from a full 3D grid of the nonlinear density field.  There is a significant difference between these two, as the former version of reconstruction is a highly under-constrained problem -- there are many more modes that shape the galaxy field than there are observed galaxies.  This aspect shapes the characteristics of the reconstructions presented here.  (The best algorithms applied to the full density field are able to reconstruct wavenumbers that are $2-3\times$ higher to $k\sim 1~$Mpc$^{-1}$ \citep{2020arXiv201200240A}.  The limit of full density field algorithms is likely set instead by where information is erased by nonlinear evolution, such as shell crossings.) 
  
   Nonlinear galaxy reconstruction algorithms generically find that they are able to reconstruct the large-scale modes in a manner that appears to be roughly limited by shot noise at low wavenumbers.  Above some wavenumber, their efficacy falls off a cliff, with studies finding that over a factor of $\sim 2$ in wavenumber the reconstructed field goes from highly correlated to uncorrelated with the input field \citep{yu17, modi18}.  We aim to understand the principles that shape this seemingly generic behavior.  Another unresolved issue is whether shot noise sets the floor for how well low wavenumber modes in the galaxy field can be reconstructed.   This issue is of high import for detecting the large-scale signatures of primordial non-gaussianity \citep{2009PhRvL.102b1302S} and neutrino mass \citep{2016PhRvD..93j3526L}.  Studies have shown that galaxy surveys, when weighting by halo mass, can have effective noises that are substantially smaller than the naive number-weighting shot noise estimate, but with a character that is still shot noise-like \citep{2009PhRvL.103i1303S, 2010PhRvD..82d3515H, 2018arXiv181110640S}.  However,  there is no understanding beyond that derived from brute-force numerics of the degree to which shot noise can be avoided.  
   We present results that suggest that at low wavenumbers it may be possible to evade shot noise by a larger factor than has yet been achieved.

Density field reconstruction addresses perhaps the deepest conceptual issue in large-scale structure --  the ultimate limit for reconstructing primordial information from a galaxy survey.  Because the methods for reconstructing the density field are so complex and computational expensive, they have not afforded a conceptual understanding of what sets this limit, even in a simplified setting that ignores the additional complexities of redshift space distortions and baryonic physics.  We also ignore these complicating factors here. 
   We attempt to make traction by understanding reconstruction in a toy universe in which galaxies of halo mass $M$ reside at the peaks of a Gaussian random linear density field, with the condition that these peaks exceed the collapse threshold for a spherical system when the field is smoothed on the Lagrangian mass scale $M$.  These peaks are then displaced with linear order Lagrangian perturbation theory (the Zeldovich approximation).  This setup is motivated by the successes of (1) excursion set theory in explaining the halo mass function \cite[e.g.][]{bond91, 2002PhR...372....1C} and (2) of Lagrangian perturbation theory \cite{1970A&A.....5...84Z, 2012JCAP...10..006T, 2014MNRAS.439.3630W}.  We highlight the visualizations in \cite{2012JCAP...10..006T}, which show that the Zeldovich approximation fares excellently at describing particle displacements, erring primarily on virialized scales.  


This paper is set up as follows.  Section~\ref{sec:motivations} discusses the scales involved in reconstruction, highlighting a curious coincidence that further motivates this work.  Section~\ref{sec:toyuniverse} presents the toy problem that we aim to solve, as well as a linear simplification. 
   Section~\ref{sec:linearmodel} studies in detail the linear model, a setup that has bearing on understanding the limits of reconstruction.  Lastly, Section~\ref{sec:nonlinearmodel} considers our full nonlinear model. 
   
    Throughout we adopt the discrete Fourier convention in a cubic volume $V$ that is common in cosmology with inverse transform given by $F(\bfx) = V^{-1}\sum_{\forall \bfk} \widetilde{F}(\bfk) \exp[-i \bfk \bfx]$, where the $\bfk$ cover a 3D grid with spacing $2\pi/V^{1/3}$, the sum runs over both positive and negative values of each component of the wavevector, and the tilde denotes the Fourier dual.  All of our calculations consider $z=0.5$ and, unless otherwise specified, are in the concordance $\Lambda$CDM cosmology with $\Omega_m=0.3$, $\Omega_\Lambda=0.7$, $n_s=0.96$, $\sigma_8=0.8$, and $\Omega_b =0.045$.


%

\section{Characteristic wavenumbers that shape reconstruction}
\label{sec:motivations}

\begin{figure}
\begin{center}
\epsfig{file=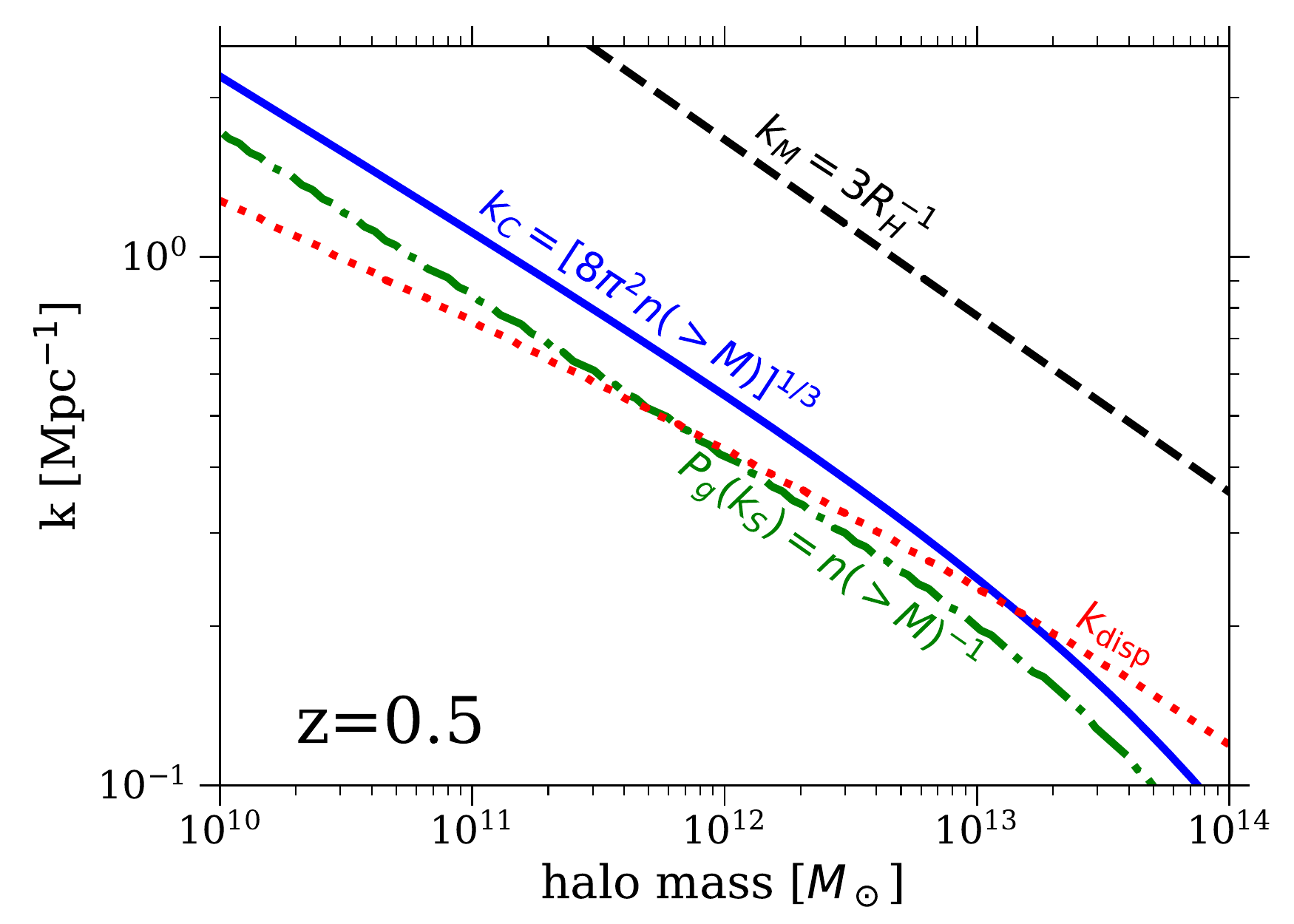, width=7.6cm}
\epsfig{file=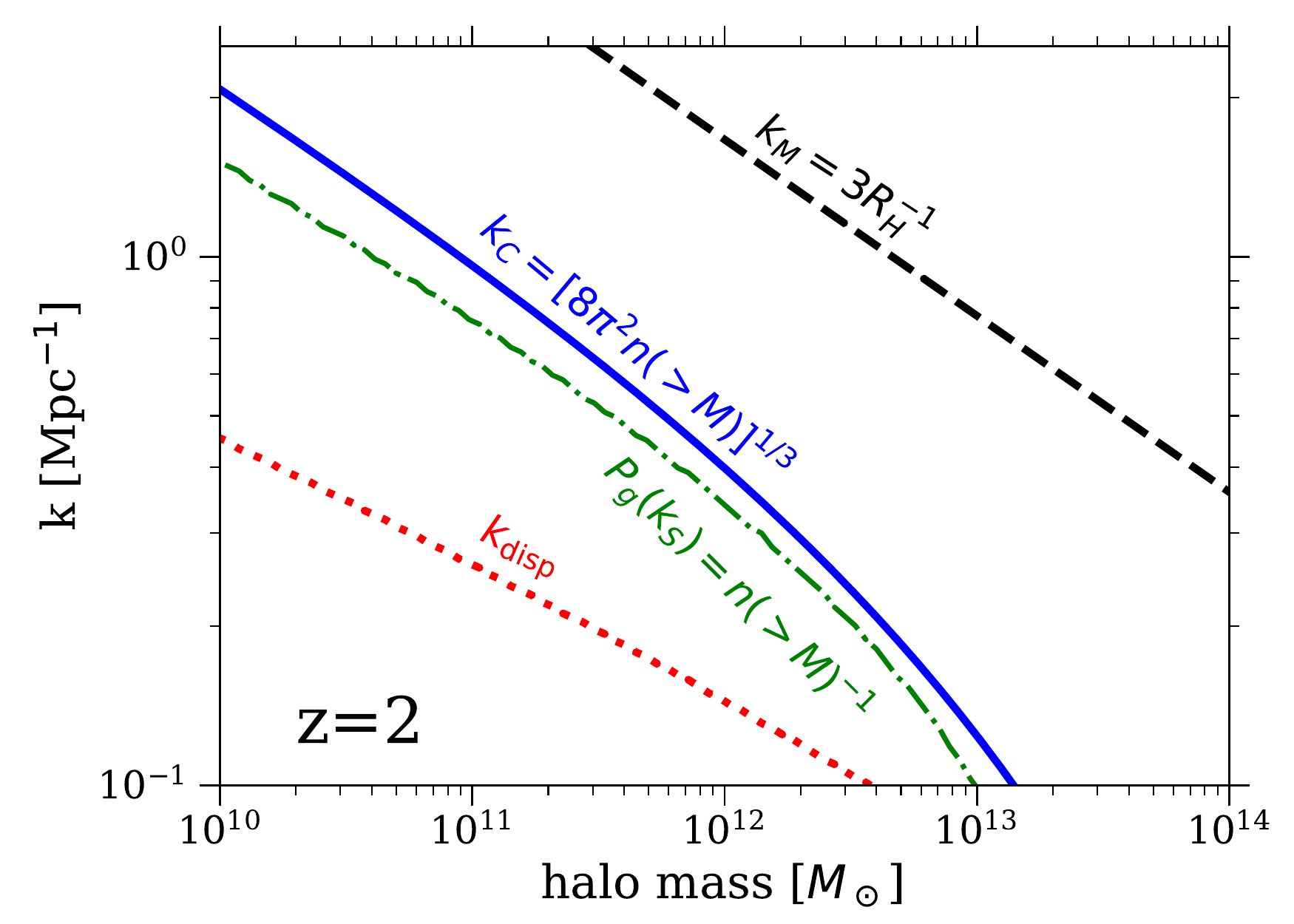, width=7.6cm}
\end{center}
\vspace{-0.5cm}
\caption{Visualization of the characteristic wavenumbers in the fiducial $\Lambda$CDM cosmology at $z=0.5$ (left) and $z=2$ (right).  Four characteristic wavenumbers are shown as a function of halo mass: (1)  the maximum wavenumber that contributes to the overdensity at the halo scale,  $k_M = 3 R_{{\rm H}}^{-1}$, where $R_{{\rm H}}$ is the Lagrangian radius of a halo (black long dashed curves), (2) the wavenumber where shot noise power is equal to the linear clustering power for the specified redshifts, $k_S$ (green dot-dashed curves), (3) the wavenumber for which the number of modes with smaller values equals ${\cal N} =4$ times the number of galaxies, $k_C$ (blue solid curves), and (4)  the wavenumber above which the displacement is the Lagrangian size of a halo, $k_{\rm disp}$ (red dotted curves).  Cumulative halo number densities, $n(>M)$, are computed using the mass function of \cite{2001MNRAS.321..372J} and assume $100\%$ completeness above the specified halo mass.  A wavenumber that is not shown is $k_{\rm NL}$, which at $z=0.5-2$ falls in the ballpark of these other wavenumbers with $k_{\rm NL} \sim 0.2-0.5$Mpc$^{-1}$.  Large-scale structure theory always considers $k_S$ and never $k_C$, but, interestingly, $k_S$ and $k_C$ are very similar in the concordance cosmology at all relevant redshifts and halo masses, perhaps explaining why $k_C$ is never considered.  We argue in \S~\ref{sec:nonlinearmodel} that nonlinear reconstruction becomes a convex problem to the extent $k_{\rm disp}$ is smaller than the other characteristic wavenumbers.    \label{fig:scales}}
\end{figure}

We are aware of several characteristic wavenumbers that may affect our ability to extract the linear modes from spectroscopic galaxy observations.   Curiously, in our universe they all fall within a factor of a few of one another at low redshifts and for motivated galaxy number densities.  The first is the nonlinear wavenumber $k_{\rm NL}$, although it enters the least frequently in this paper's discussion.  We define it as the solution to 
\begin{equation}
\Delta_{L}(k)^2 \equiv \frac{k^3}{2\pi^2} D(z)^2 P_{L,0}(k)  = 1.
\label{eqn:knonlinear}
\end{equation}
  Here $P_{L,0}$ is the $z=0$ power spectrum of the linear matter overdensity, and $D(z)$ is the linear growth factor normalized to a present value of unity (such that the linear power at any redshift is $P_{L} = D(z)^2 P_{L,0}$). Solving for this scale in the concordance $\Lambda$CDM cosmology yields $k_{\rm NL} = 0.2$~Mpc$^{-1}$ at $z=0.2$ and  $k_{\rm NL} = 0.5$~Mpc$^{-1}$ at $z=1.0$. The nonlinear wavenumber crudely bounds where perturbation theory is applicable.\footnote{Formally, the density variance -- which defines our nonlinear scale -- is not the only parameter that shapes nonlinearity in large-scale structure in the perturbative limit.  The linear displacement contributed by large and small scales relative to the wavenumber in question are additional ordering parameters \citep[e.g.][]{2015JCAP...02..013S}. 
 The latter nonlinear parameter does enter into our discussion, setting the characteristic wavenumber $k_{\rm disp}$ (eqn.~\ref{eqn:sigmadisp}).}  
 Around $k_{\rm NL}$, we can approximate the dimensionless linear matter overdensity power spectrum as
\begin{equation}
\Delta_L(k)^2 =\left(\frac{k}{k_{\rm NL}}\right)^{3-n_{\rm eff}},
\end{equation}
where $n_{\rm eff}$ is an effective power-law index.   For our concordance $\Lambda$CDM cosmological model, $n_{\rm eff} \approx 2$ at $k\sim 0.1-1~$Mpc$^{-1}$, wavenumbers where reconstruction can be effective. 

In addition to the nonlinear wavenumber, there is the wavenumber where the shot noise power is equal to the linear clustering power:
\begin{eqnarray}
b_L^2 D(z)^2 P_{L,0}(k_S) = \bar n^{-1} &\overbrace{\xrightarrow{\hspace*{.8cm}}}^{\text{power law}}& ~~  k_S = \left(2\pi^2 b_L^2 \bar n k_{\rm NL}^{n_{\rm eff} - 3} \right)^{1/n_{\rm eff}}, \\
&\overbrace{\xrightarrow{\hspace*{.8cm}}}^{\rm \Lambda CDM}&~~ k_S \approx 0.4 {\rm ~Mpc}^{-1} ~\left (b_L^2  D^2  \bar n_{-3} \right)^{0.5},
\end{eqnarray}
where $\bar n_{-3} = \bar n/[10^{-3} \Mpc^{-3}]$, $\bar n$ is the galaxy number density, and $b_L$ is the galaxy linear bias.  

Additionally, there is another characteristic wavenumber that is not mentioned in discussions of reconstruction but that we argue is relevant -- the ``constraints wavenumber'', $ k_C$.  We define $k_C$ as the wavenumber where the number of modes with smaller wavenumbers is equal to the number of constraints, which in the large-volume limit is solved by:
\begin{equation}
{\cal N} \bar n V =  \frac{V k_C^3}{2\pi^2}  ~~~~~{\xrightarrow{\hspace*{.8cm}}} ~~~~~ k_C = 0.4 \left(\frac{{\cal N}}{4} \right)^{1/3} n_{-3}^{1/3}.
\end{equation}
We will motivate ${\cal N} \approx 3$ constraints per galaxy, owing to the three positions for each galaxy, and sometimes ${\cal N} \approx 4$ when the galaxies' halo masses can be precisely estimated.  In a picture where reconstruction is able to constrain large-scale modes before small-scale ones, $k_C$ should bound the wavenumbers that can be reconstructed.  
 
Another characteristic wavenumber is set by the Lagrangian size of halos, which approximates the maximum wavenumber that influences halo formation.  We define this wavenumber to be 
\begin{equation}
k_{M} = 3 R_{{\rm H}}^{-1} =  1.7~ {\rm Mpc}^{-1}~ \left( \frac{M}{10^{12} M_\odot} \right)^{-1/3},
\label{eqn:kM}
\end{equation}
where $M$ is the halo mass and $R_{{\rm H}} \equiv [3M/(4\pi \rho_m)]^{1/3}$.  At $k_M$, the Fourier dual of a real-space tophat window function with unit support is equal to $\approx 1/3$.  
 
 Lastly, there is the scale where the displacement from wavenumbers greater than $k_{\rm disp}$ is larger than the halo Lagrangian radius, $R_{{\rm H}}$.  We estimate this wavenumber by using that the aligned pairwise variance of the Zeldovich Approximation displacement that is contributed by modes with wavenumbers greater than $k$,
 \beq
 \sigma_{\rm \Psi}(k)^2 =  \int_{k}^\infty \frac{dk'}{6\pi^2} P_L(k'),
 \label{eqn:sigmadisp}
 \eeq
  and then solving $\sigma_{\rm \Psi}(k_{\rm disp}) = R_{{\rm H}}$ for $k_{\rm disp}$.   We will show that to the extent that this wavenumber is much smaller than the previous characteristic wavenumbers, the modes that determine the position of a halo may be reconstructed well enough to make the nonlinear reconstruction problem approximately convex (\S~\ref{sec:nonlinearmodel}).

It is notable that in the concordance cosmology all of these characteristic wavenumbers are within a factor of several of one another.  Figure~\ref{fig:scales} shows these wavenumbers as a function of the minimum halo mass surveyed, assuming 100\% completeness so that all halos above the specified mass are included in the survey weighted by number density.   The left panel is for $z=0.5$, and the right for $z=2$.  Intriguingly, the constraints scale $k_C$ and the shot noise scale $k_S$ are nearly the same at all halo masses. This similarity is a coincidence of our cosmology, as these characteristic wavenumbers would scale differently with $M$ in cosmologies with different effective scalar spectral indices ($n_{\rm eff}$).  That $k_C(M) \approx k_S(M)$ in the concordance cosmology may explain why $k_C$ never is referenced in large-scale structure literature, even though we will show it sets the maximum wavenumber that is accurately reconstructed.  That $k_{M}$ is a factor of $\approx 5$ larger than $k_C$ across all halo masses indicates that reconstruction is always under-constrained, with $\sim 5^3$ more modes that shape the halo field than constraints.  This under-constrained property of reconstruction shapes many of the results in this study.

\section{Reconstruction in a toy universe}
\label{sec:toyuniverse}

Our aim is to reconstruct the initial conditions from 3D galaxy position measurements.   This paper focuses on a toy model for the cosmos that is more tractable for understanding the limits of reconstruction, while still maintaining significant similarities to our universe.  The model places galaxies with halo mass $M$ at peaks in the smoothed-on-scale-$M$ linear overdensity field that meet some threshold overdensity, $\delta_c(M)$.\footnote{Our approach could be straightforwardly generalized to more realistic models that consider eigenvalues of the local tidal field \citep[e.g.][]{1993MNRAS.265..689K, 1996ApJS..103....1B}.}  We then displace these peaks with linear order Lagrangian perturbation theory (the Zeldovich approximation).  This setup is motivated by the successes of (1) excursion set theory in explaining the halo mass function \citep[e.g.][]{bond91, 2002PhR...372....1C} and (2)  Lagrangian perturbation theory \citep{1970A&A.....5...84Z, 2012JCAP...10..006T, 2014MNRAS.439.3630W}.  

Let us assume a survey with cubic volume $V$ with a list of galaxies (i.e. overdensity peaks) with positions $\bfx_j$ and masses $M_j$ for $j \in [0,N)$.  The condition for peak heights reduces reconstruction to $N$ constraint equations: 
\beq
 V^{-1} \sum_{\forall  \bfk} \tdelta_{\bfk} e^{-i \bfk \cdot \left[ \bfx_j - \boldsymbol{\psi}(\bfq_j |\tdelta_{\bfk'}) \right]} W_{M_j}(k) = \delta_c(M_j)  ~~~~  \text{for  $j \in [0,N)$},
\label{eqn:master}
\eeq
where $\bfq_j$ is the Lagrangian position of the $j^{\rm th}$ halo and $W_M(k)$ is a window function that approximates the Lagrangian size of the halo (which hosts the `observed' galaxy), here taken to be a tophat in real space, and $\boldsymbol{\psi}$ is the Zeldovich approximation displacement vector that is given by\footnote{It might be more realistic to also include the halo window function $W_M(k)$ in the integral in eqn.~\ref{eqn:disp} for $\boldsymbol{\psi}$, as modes with wavelengths smaller than the halo scale do not contribute significantly to its displacement.  However, we find that none of our results are appreciably changed if we include this factor.  One could also add to the displacement vector an effective term $\propto k \tdelta_\bfk$ owing to small-scale dynamics \cite{baumann12, carrasco12}, but we suspect it also will have limited effect. }
\beq
\boldsymbol{\psi}(\bfq_j | \tdelta_\bfk) = -i V^{-1} \sum_{\forall \bfk}   \frac{\bfk}{k^2} \; \tdelta_\bfk e^{-i \bfk \cdot \bfq_j}.
\label{eqn:disp}
\eeq
We want to use the constraints given in eqn.~\ref{eqn:master} (plus possibly additional constraints discussed shortly) to reconstruct the \emph{linear theory} modes $\tdelta_\bfk$.   As emphasized in \S~\ref{sec:motivations}, if we take the maximum wavenumber that contributes to the formation of halos for masses typical of modern spectroscopic galaxy surveys (the wavenumber ``cutoff'' in $W_M$), this is a highly under-constrained problem with (infinitely) many solutions.  We aim to select a solution that is as close as possible to the input density field. 

One difficulty with the above setup is that $\boldsymbol{\psi}$ is a function of the Lagrangian coordinate of each halo $\bfq_j$, which is not an observable.  One can avoid this difficulty by substituting in the Taylor expansion $ \boldsymbol{\psi}(\bfq) =  \boldsymbol{\psi}(\bfx) -   (\boldsymbol{\psi}(\bfx) \cdot \nabla  )  \boldsymbol{\psi}(\bfx) +... $ or by some other optimization technique.  However, we instead make the simplifying assumption that the displacement evaluated at the Lagrangian position of a galaxy can be calculated from $ \boldsymbol{\psi}$ and the final position of the galaxy $\bfx_j$.  We expect that the uncertainty in this mapping will not be what sets the efficacy of reconstruction.\footnote{Our results appear  insensitive to $\sim 2~$Mpc errors in the displacement that would occur from the laziest (zeroth order) approximation $\boldsymbol{\psi}(\bfq_j) \approx  \boldsymbol{\psi}(\bfx_j)$.}

With this simplification, it is useful to instead formulate our equations in Lagrangian space.  The master set of equations we are considering becomes
\begin{eqnarray}
 &V^{-1} \sum_{\forall  \bfk} \tdelta_{\bfk} e^{-i \bfk \cdot \left[ \bfq_j -  \Delta \boldsymbol{\psi}(\bfq_j |\tdelta_{\bfk'}) \right]} W_{M_j}(k) = \delta_c(M_j) ~~~~  \text{for  $j \in [0,N)$},
\label{eqn:constraintsnonlinear}
\end{eqnarray}
where $\Delta  \boldsymbol{\psi} \equiv \boldsymbol{\psi} - \boldsymbol{\psi}_{\rm TRUTH}$.
Here $\boldsymbol{\psi}_{\rm TRUTH}$ is the \emph{true} displacement computed from the input $\tdelta_{\bfk}$ -- i.e. the field we aim to reconstruct.  One might worry that now we have written the problem in terms of some unobservable quantities ($\bfq_j$, $\boldsymbol{\psi}_{\rm TRUTH}$), rather than our observables $(\bfx_j, M_j)$.  However, this is a sleight of hand as  $\bfq_j = \bfx_j - \boldsymbol{\psi}_{\rm TRUTH}(\bfq_ j)$, noting also the simplification in the previous paragraph.

We can identify \emph{linear} equations that, if satisfied, also solve Eqn.~\ref{eqn:constraintsnonlinear}, namely for $j \in [0,N)$:
\begin{eqnarray}
\delta_c(M_j) - V^{-1} \sum_{\forall \bfk} \tdelta_{\bfk} e^{-i \bfk \cdot \bfq_j} W_{M_j}(k) &=& 0 ;
\label{eqn:lagrangian}\\
\boldsymbol{\psi} (\bfq_j | \tdelta_\bfk)- \boldsymbol{\psi}_{\rm TRUTH}(\bfq_j)&=&0.
\label{eqn:displacements}
\end{eqnarray}
These are the conditions that the density smoothed on the scale $W_{M_j}$ adds up to the collapse threshold $\delta_c(M_j)$ at the Lagrangian position $\bfq_j$ and that the displacement is the true displacement.  They embody more information than an observer is able to access as they require knowledge of the $\bfq_j$ and hence $\boldsymbol{\psi}_{\rm TRUTH}(\bfq_j)$.  Solutions to these conditions are not the only solutions to our nonlinear equations (Eqn.~\ref{eqn:constraintsnonlinear}), but solutions to these linear equations clearly would be preferred ones.   Despite this linear system of equations also being extremely under-constrained (as these $4$N equations should be far fewer than the total number of modes), these equations' linearity in $\tdelta_\bfk$ means that one can always find a solution, whereas gradient descent-like methods for solving the nonlinear equations may easily get stuck in local minima (\S~\ref{sec:nonlinear}).   

``Reconstruction'' from the linear eqn.s \ref{eqn:lagrangian} and \ref{eqn:displacements} will be more successful than efforts starting with the nonlinear master equation (eqn.~\ref{eqn:master}) as these equations require $\boldsymbol{\psi}_{\rm TRUTH}(\bfq_j)$, which a lowly cosmologist doing reconstruction would have no way of knowing exactly.  Therefore, ``reconstruction'' using these linear constraint equations bounds the effectiveness of reconstruction.  We later obtain solutions to our nonlinear equations that come close to saturating the bounds placed by solving the linear equations.  Furthermore, we show that other reconstruction algorithms that have been applied to the density field in N-body simulations \cite{yu17, modi18} come close to saturating these bounds.\\   


One might worry that our idealized setup does not capture all the information used by nonlinear reconstruction algorithms, which are trying to model the full halo distribution.   For example, the setup outlined so far does not account for halos forming at density peaks in Lagrangian space: 
\beq
 - i V^{-1} \sum_{\forall \bfk} \bfk  \tdelta_{\bfk} e^{-i \bfk \cdot  \left[ \bfq_j -  \Delta \boldsymbol{\psi}(\bfq_j |\tdelta_{\bfk'}) \right]} W_{M_j}(k) = 0 ~~~~  \text{for  $j \in [0,N)$}.
\label{eqn:peaks}
\eeq
Additionally, there may be other constraints on the shape of the surrounding Lagrangian overdensity, like
\beq
 -V^{-1} \sum_{\forall  \bfk} k^2 \tdelta_{\bfk} e^{-i \bfk \cdot  \left[ \bfq_j -  \Delta \boldsymbol{\psi}(\bfq_j |\tdelta_{\bfk'}) \right]} W_{M_j}(k)  =  -\ell_j^{-2}\delta_c(M_j)  ~~~~  \text{for  $j \in [0,N)$},
\label{eqn:peaksize}
\eeq
such that, for example, if the characteristic size is $\ell_j \gg R_j \equiv (3M/4\pi)^{1/3} $ the halo is likely quickly growing in mass.  
  Because the $\tdelta_\bfk$ in these additional constraint equations are weighted by powers of $k$, they are less important for constraining the low wavenumber modes that can be reconstructed accurately. 
  Indeed, we demonstrate in \S~\ref{ss:excursionset} that adding the three peak constraints per galaxy, eqn.~(\ref{eqn:peaks}), results in a modest improvement in the reconstructed field.  
   Lastly, if our galaxy survey is complete for halos above a given mass, then there should be constraints enforcing that there are not other peaks on this mass scale.  However, in practice the solutions we find tend not to include many additional peaks, and so this condition is more or less satisfied naturally.  

\subsection{Conditioning to select a solution and solving (the linearized toy problem)}
\label{ss:conditioning}
Our under-constrained system of equations can be cast as a least squares optimization problem.  \emph{Specializing first to the linear set of conditions}, which require godlike knowledge of $\boldsymbol{\psi}_{\rm TRUTH}(\bfq_j)$, we search for the global minimum of some or all of the terms in the following loss function:
\begin{eqnarray}
{L} =&& \sum_{j=0}^{N} \Bigg[ \overbrace{\left(\delta_c(M_j) - V^{-1} \sum_{\forall \bfk} \tdelta_{\bfk} e^{-i \bfk \cdot \bfq_j} W_{M_j}(k) \right)^2}^{\text{Lagrangian overdensity (L)}}  +
\overbrace{\ell_{\rm D}^{-2}\left( \boldsymbol{\psi} (\bfq_j | \tdelta_\bfk)- \boldsymbol{\psi}_{{\rm TRUTH}}(\bfq_j) \right)^2}^{\text{displacements (D)}}
   \nonumber\\
&+&  \underbrace{\ell_{\rm P}^{2} \left(V^{-1}   \sum_{\forall \bfk} \bfk \tdelta_{\bfk} e^{-i \bfk \cdot \bfq_j} W_{M_j}(k) \right)^2}_{\text{peaks (P)}} \Bigg]  +\underbrace{C(\tdelta_\bfk)}_{\text{regularization}}.
\label{eqn:linear}
\end{eqnarray}
 We later consider the nonlinear case in a very similar setup. This problem has infinitely many minima with $L=0$ if $C=0$, since the number of modes that shape the density field is much greater than the number of constraint equations ($7N$ if we include all three conditions in eqn.~\ref{eqn:linear}).  The regularizer $C$ has to supplement with sufficiently many conditions to constrain the system.  Furthermore, $C$ has to be the square of an expression linear in $\tdelta_\bfk$ to retain the desirable property of $L$ being positive and solvable with linear algebra (this choice is called Tikhonov regularization; \citep{nr}).  
  Fortuitously, the Gaussian mode-amplitude prior that reconstruction algorithms employ \citep[e.g.][]{2017JCAP...12..009S} fall exactly in this category, generalizing in our noiseless case to the Ridge Regression regularization condition of
\begin{equation}
C(\tdelta_\bfk) = \sigma^2 \sum_{\forall \bfk}  \frac{|\delta_\bfk|^2}{P_L(k) V},
\label{eqn:CG}
\end{equation}
which is a special case of Tikhonov regularization.  While previous reconstruction algorithms are phrased in terms of maximizing a posterior rather than a `noiseless' loss function, the loss function is analogous to the logarithm of the posterior when divided by $2\sigma^2$.  The reconstruction ``noise'' in our case (and that we argue holds for posterior reconstruction algorithms) is not shaped significantly by the ``error'' parameter $\sigma$, but rather by how modes project onto the set of several times $ N$ well-constrained quantities.  Modes that are not constrained are set to zero by this regularization.  Indeed, we show later that our solution to linear equations depends negligibly on choice of $\sigma$ once $\sigma^2\lesssim 0.1$, which can be recast in terms of the error on e.g. the displacements for modes to be constrained at a cosmologically interesting level.  Finally, the regularization does assume a cosmology to calculate the linear power spectrum, $P_L(k)$ (and we use here the $P_L(k)$ of our background cosmology), but our results are relatively insensitive to this assumption for reasons that will be discussed. 

One issue with using the Gaussian regularization (eqn.~\ref{eqn:CG}) is that when a mode is not well constrained, the solution that minimizes ${L}$ is  $\tdelta_\bfk = 0$.  Thus, the global minimum will be a biased solution that favors lower power in less constrained pixels.  This is not a surprise and is analogous to the down-weighting of noisy pixels in optimal map making (Weiner filtering).  We have also investigated a Tikhonov regularization condition that preferences a particular random phase field with power spectrum given by the input (and with the same $[V P_L(k)]^{-1}$ weighting as above).  This alternative setup retains the property of linearity, but conditions to a non-Gaussian field.  We find that the wavenumbers that can be reconstructed are not significantly changed in this alternative regularization scheme.
 


In addition to the regularization condition, we need to choose some physically motivated values for the scales that appear in $L$, namely $\ell_{\rm D}$ and $\ell_{\rm P}$, as these will weight the equations in different manners.  We take $\ell_{\rm D}=\ell_{\rm P} = 10~$Mpc for our fiducial values as motivated in footnote~\ref{foot5}.\footnote{\label{foot5}We can weakly motivate values by expanding eqn.~\ref{eqn:constraintsnonlinearexpand} in  $\Delta \boldsymbol{\psi}$ such that
\begin{equation}
 V^{-1} \sum_{\forall  \bfk} \tdelta_{\bfk} e^{-i \bfk \cdot \bfq_j}W_{M_j}(k)  \left[1 - i \bfk  \cdot \Delta \boldsymbol{\psi} - \frac{1}{2}(\bfk \cdot \Delta \boldsymbol{\psi})^2 +\ldots \right]  = \delta_c(M_j) ~~~~  \text{for  $j \in [0,N)$},
\label{eqn:constraintsnonlinearexpand}
\end{equation}
In the square brackets, the different terms bear semblance to the Lagrangian overdensity, the peak condition, and the peak shape condition.  Defining $\widehat {\delta_{c}}(\bfq, M_j) \equiv V^{-1} \sum_{\forall  \bfk} \tdelta_{\bfk} e^{-i \bfk \cdot \bfq}W_{M_j}(k)$ and ${\ell_j}$ describes the $j^{\rm th}$ peak's radius of curvature (eqn.~\ref{eqn:peaksize}), then
\begin{equation}
\left(\widehat \delta_c(\bfq_j, M_j)  - \delta_c(M_j) \right) +  \boldsymbol{\Delta \psi} \cdot \nabla_{\bfq}  \widehat{\delta_c}(\bfq, M_j)\Big |_{\bfq_j}  -  \frac{\delta_c(M_j)}{6 {\ell_j}^{2}} \boldsymbol{\Delta \psi} ^2  +\ldots    = 0 ~~~~  \text{for  $j \in [0,N)$},
\label{eqn:motivateell}
\end{equation}
assuming that the peak curvature and displacement are uncorrelated to deduce the last term. 
 Eqn.~\ref{eqn:motivateell} is \emph{somewhat} analogous to our loss function ($L$) but where the terms related to the Lagrangian density, peaks and displacements have physically motivated coefficients rather than the parameters $\ell_{\rm D}$ and $\ell_{\rm P}$.  First, we anticipate that the curvature radius at peak, $\ell_j$, is likely somewhat larger than the Lagrangian halo size or $R_{{\rm H}} \sim 3~$Mpc for halos in the mass range we consider.  For simplicity, and because of the factor of $6$ in the denominator, we set $\ell_{\rm D}= 10~$Mpc.  We seldom consider the peaks constraint, nor does equation~\ref{eqn:motivateell} motivate a clear choice for $\ell_{\rm P}$.}  We have decrease/increased $\ell_{\rm D}$ and $\ell_{\rm P}$ by a factor of 5 and find that the cross correlation coefficient of our solution with the truth is negligibly altered.  
  This lack of dependence we think owes to the under-constrained nature of problem:  there are solutions that satisfy all our constraints perfectly in the limit that the regularizer normalization $\sigma \rightarrow 1$  (\S~\ref{ss:regularization}).  When we turn to the nonlinear problem in \S~\ref{sec:nonlinearmodel},  $\ell_{\rm D}$ and $\ell_{\rm P}$ no longer are relevant.

\subsection{A numerical realization of a simplified universe}
\label{ss:excursionset}
We now design a realization of the universe that embodies our toy picture of a critical overdensity threshold in Lagrangian space plus its Zeldovich approximation displacement.  We further desire the spectrum of halo masses to match the halo mass function found in N-body simulations.  To achieve the latter, we implement a model that is motivated by extended Press-Schechter theory \citep{bond91}, except that it also attempts to capture halo discreteness.  This algorithm is most analogous to the `peak patch' scheme for generating a galaxy field of \citep{1996ApJS..103....1B, 2020arXiv200108787S}.  

To create the linear matter overdensity field, we generate a Gaussian random field with power spectrum $P_L(k)$ in a cube of volume $V$ with $N_s^3$ discrete samples on a grid.  We then smooth the overdensity field on different scales with a tophat in real space specified by its enclosed Lagrangian mass $M$, starting with the largest mass scale and moving to smaller and smaller masses.  For the field filtered at each smoothing scale $\delta_M$, we identify all the peaks that satisfy $\delta_M(z) > \delta_c$, where $\delta_c = 1.7$ as motivated by calculations in spherical collapse \citep{1972ApJ...176....1G}.  Starting with the highest peak, we mark off a spherical Lagrangian region of mass $M$; all cells that fall within this region can no longer be used as the center of a halo.  We further associate the $j^{\rm th}$ peak with a halo specified by its Lagrangian position $q_j$ and mass $M_j$.  The density field is then smoothed over a smaller scale, and the process is repeated, heeding the prior exclusions.  The algorithm results in a list of `halo' positions, masses, and peak overdensities.  To generate the real space position of a halo $\bfx_j$, we then displace the peak with its Zeldovich approximation displacement, $\boldsymbol{\psi}_{\rm TRUTH}(\bfq_j)$, where we remind the reader that `TRUTH' indicates the displacement is calculated with the linear overdensity field generated by this algorithm rather than the reconstructed linear field.  

\begin{figure}
\begin{center}
\epsfig{file=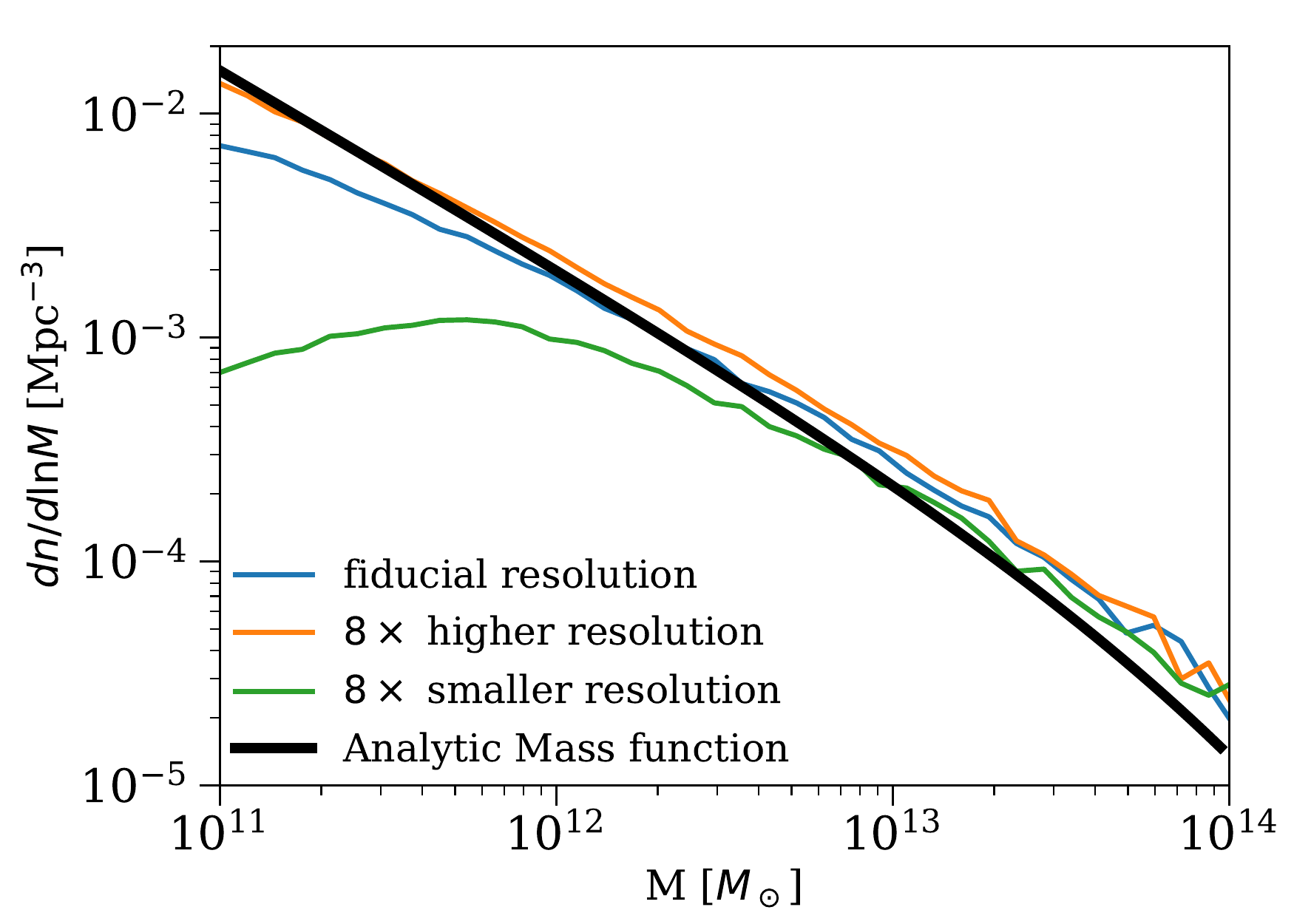, width=8cm}
\end{center}
\caption{Halo mass functions in the excursion set-inspired model used to generate this study's halo fields (\S~\ref{ss:excursionset}).  The colored curves show the mass function at different resolutions.  The blue curve is computed using the fiducial box size and grid resolution (200~Mpc, $255^3$ cells).   These halo mass functions match reasonably the fit to the mass function in cosmological N-body simulations of Jenkins et al.~\cite{2001MNRAS.321..372J} (black solid).  At the fiducial resolution, the model is well converged for halos with $M>10^{12}M_\odot$ that are our primary consideration, and it undershoots by a factor of two at the smallest halos we consider with $M=10^{11}M_\odot$.  The green curve is computed  for 200~Mpc and $127^3$ grid cells, which are the specifications used in \S~\ref{sec:nonlinear} for the more computationally expensive nonlinear model (but where we only consider halos with $M >5\times10^{12}M_\odot$).  \label{fig:massfunc}}
\end{figure}

Figure~\ref{fig:massfunc} shows the mass function that results from this algorithm.  The thick black curve is the fit to the mass function in cosmological N-body simulations of Jenkins et al.~\cite{2001MNRAS.321..372J}. 
  For the highest resolution calculations shown, our model does well at latching onto the Jenkins mass function, with a small overshoot at the largest masses.   At the `fiducial' resolution that we adopt for most of our calculations ($N_s=255$, $V^{1/3}=200~$Mpc), the mass function undershoots at $M=10^{11}M_\odot$ by almost a factor of two, but is in reasonable agreement at higher masses.  

\begin{figure}
\begin{center}
\includegraphics[totalheight=5.4cm, trim=50 2 80 10, clip=true]{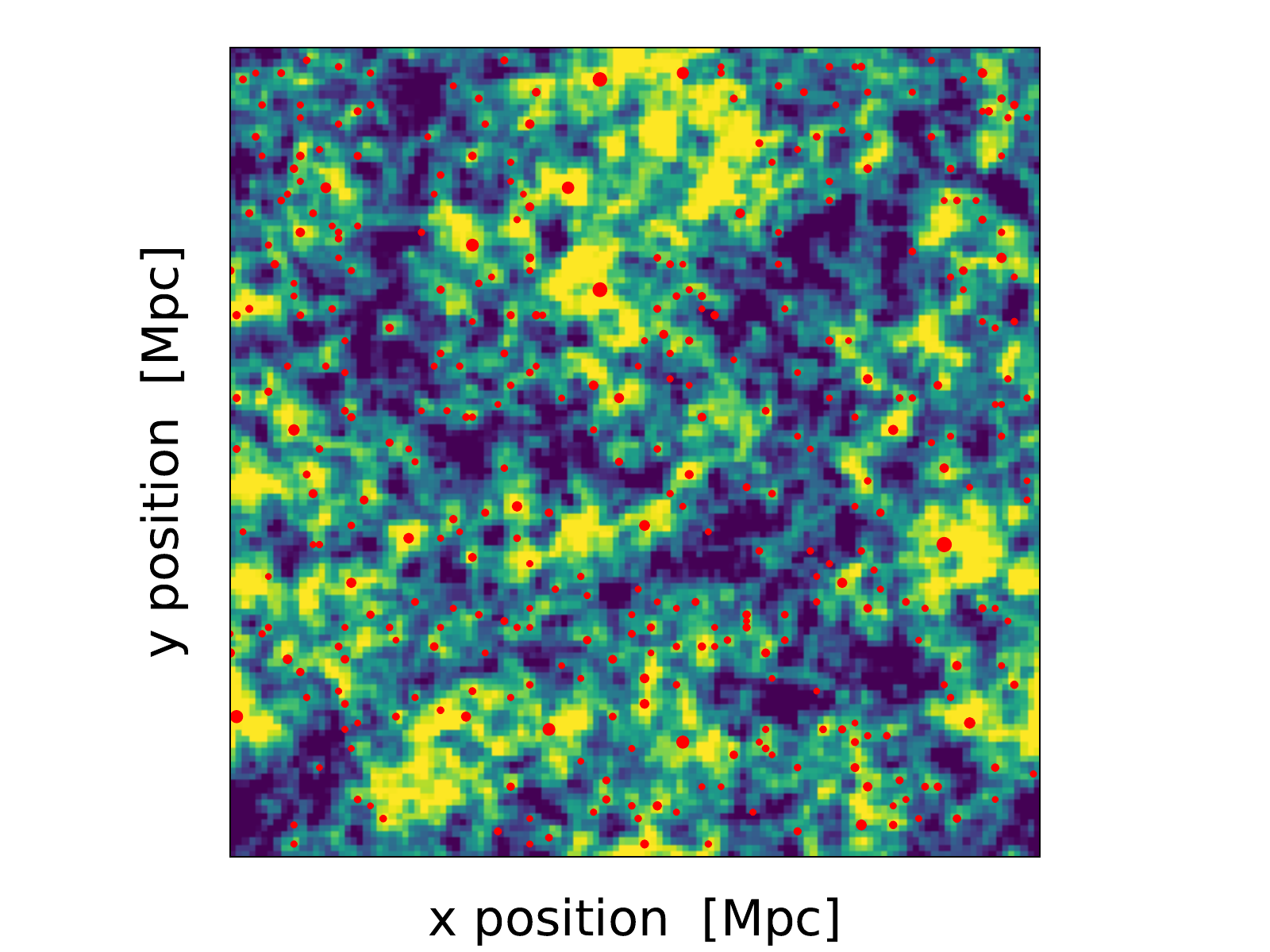}
\includegraphics[totalheight=5.4cm, trim=80 2 80 10, clip=true]{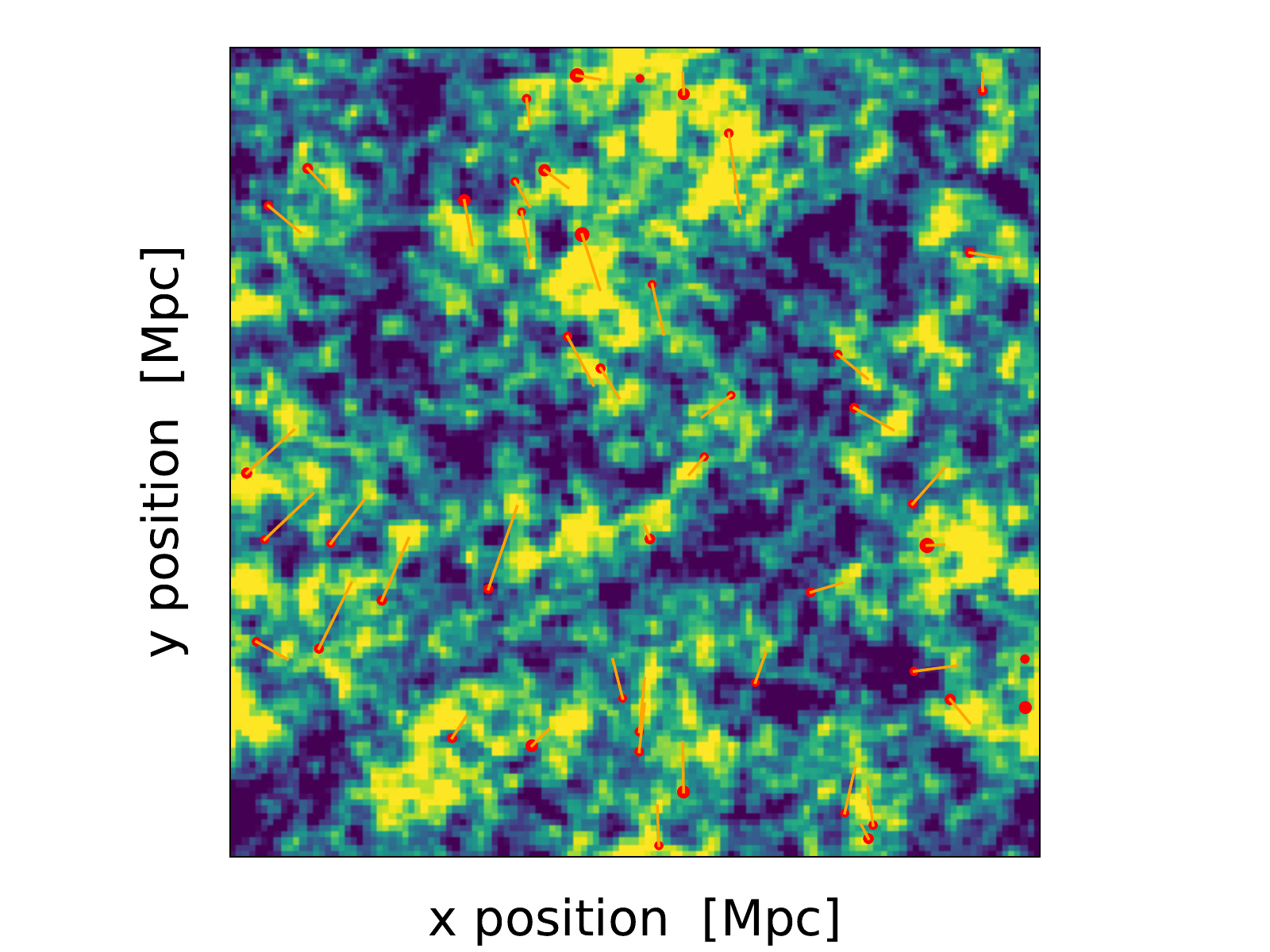}
\includegraphics[totalheight=5.4cm, trim=80 2 80 10, clip=true]{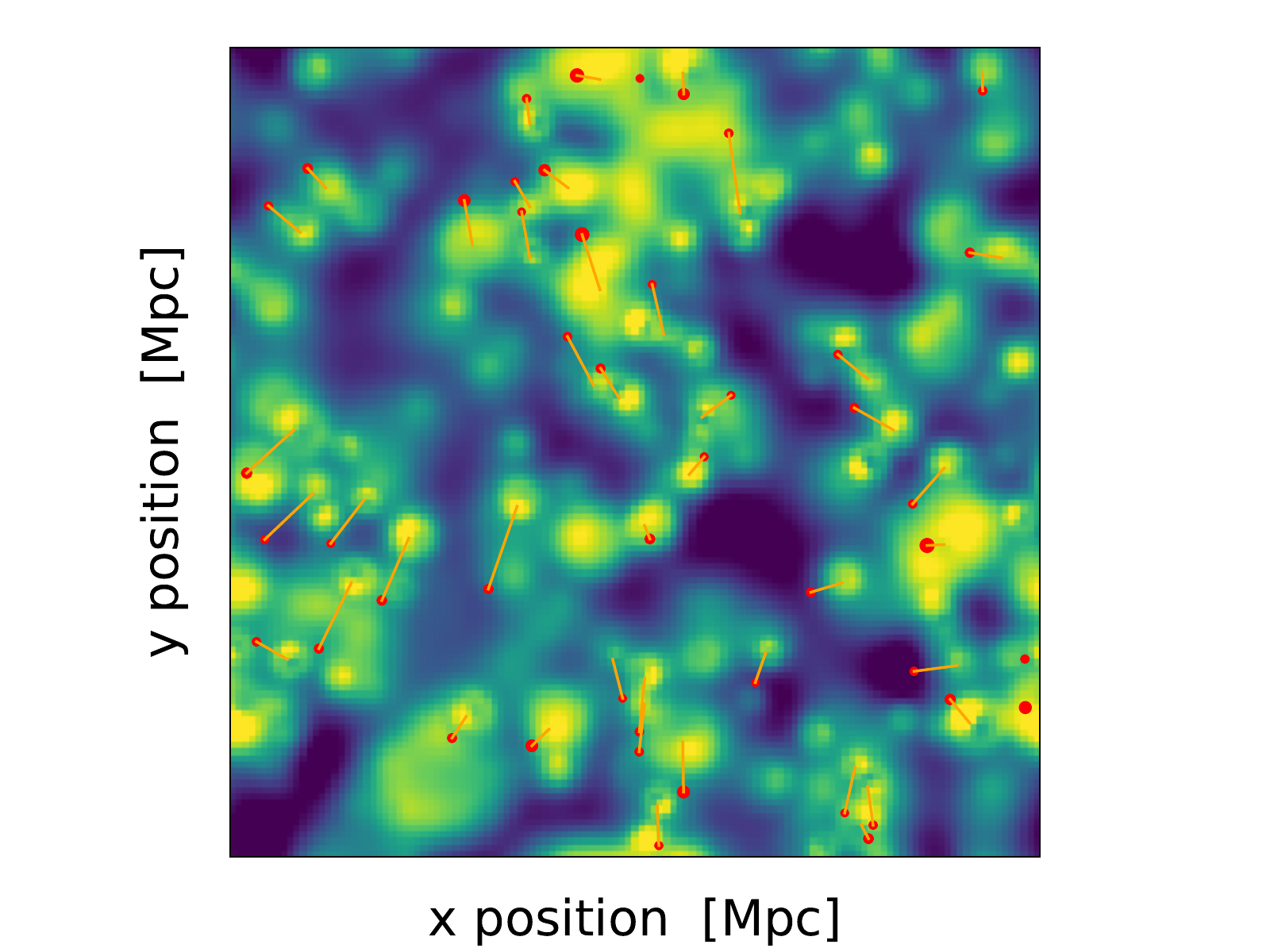}
\end{center}
\caption{\emph{The left panel} shows the Lagrangian space position of halos in our algorithm, juxtaposed on top of the linear overdensity field, for a 20~Mpc deep projection through a $100 \ $Mpc box.  The image of the linear overdensity saturates at values of [-3,3]. These calculations were performed for $127^3$ elements in a $100\;$Mpc box, essentially matching the fiducial resolution of our $200\;$Mpc box.  Halo sizes are encoded by the radius of the dot, which scales as $M^{1/3}$ above the minimum halo mass shown of $10^{11}M_\odot$.  \emph{The middle panel} shows the displaced position of halos with $M>10^{12}M_\odot$ plus the displacement vector, again juxtaposed on top of the same linear field.  Halos whose starting position is within the slice are shown (rather than selecting halos by their displaced position), and displacement vectors are not shown for halos that traverse a box boundary.  \emph{The right panel} shows our reconstruction from our linearized algorithm with all $7N$ constraint equations (the Lagrangian overdensity [L], displacements [D] and peak conditions [P]), as well as the same halo+displacement field as in the middle panel. \label{fig:algorithm}}
\end{figure}

Lastly, Figure~\ref{fig:algorithm} provides a visualization of the algorithm's output in a 20~Mpc deep projection through a $100\;$Mpc periodic box.  The image colorscale saturates at linear overdensities of [-3,3].  These calculations were performed for $127^3$ elements, matching the fiducial resolution of our $200~$Mpc box.  The left panel shows the algorithm's Lagrangian space position of halos, juxtaposed on top of the linear overdensity field. Halo sizes are encoded by the radius of the dot $r_{\rm dot}$, with $r_{\rm dot} \propto M^{1/3}$, with the left panel featuring halos with $M>10^{11}M_\odot$.  Halo exclusion around the most massive halos is evident.   The middle panel shows the displaced position of halos with $M>10^{12}M_\odot$ plus the displacement vector, again juxtaposed on top of the same linear field.  Halos whose starting position is within the slice are shown.  The righthand panel shows the results of one of our reconstruction algorithms (\S~\ref{sec:linearmodel}).

\subsection{Low memory optimization algorithms}
\label{ss:algorithms}
Our fiducial calculations attempt to constrain $255^3-1$ independent parameters from the positions of as many as $\sim 10^5$ galaxies.  If we wrote our linear optimization problem as one of matrix algebra, we would need to store a matrix of size $\sim 255^3\times 10^5$.  Fortunately, there are fast algorithms to solve such systems that do not require holding this large matrix in memory.  In particular, for the linear and nonlinear calculations presented here, we use the L-BFGS iterative algorithm, which employs a Newton's method-like algorithm to converge towards a minimum \citep{nr}, using the implementation of \url{http://www.chokkan.org/software/liblbfgs/}.  This method requires a loss function and its gradient, the latter of which can be computed analytically for our simplified model.  We terminate the iteration when the loss function has changed by less than 1\% in the prior ten iteration steps, which we find is sufficient for convergence.  (We did find that the less conservative criteria of terminating with 10\% rather than 1\% sometimes did not reach convergence at low wavenumbers for the fully nonlinear reconstruction problem.)  We have also checked that we obtain the same solution for the linear problem with the LSQR algorithm, which is an efficient conjugate gradient solver \citep{LSQR}.   

There are a few technical aspects of note.  First, rather than treating the complex mode amplitude, we perform our optimizations treating the real and imaginary components of each mode as separate parameters.  Second, we were careful to eliminate all redundancy in $\tdelta_\bfk$ that owes to it being the Fourier transform of a real field.  The reason for the non-standard grid number of $255$ in our computations rather than the more-standard $2^8= 256$ is that odd numbered grids have less redundancy in real to complex Fast Fourier transform algorithms like FFTW\footnote{\url{http://www.fftw.org}}.  The speed that we evaluate Fast Fourier transforms is not a limiting factor for obtaining a solution.  Last, we have verified that the non-conditioned loss function is zero when evaluated at the correct solution and that our analytic gradient is computed correctly by taking numerical gradients of the loss function.

\section{Reconstruction in the linearized model}
\label{sec:linearmodel}

\begin{figure}
\begin{center}
{\includegraphics[width=12.6cm]{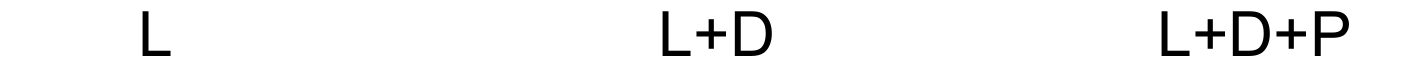}}
\includegraphics[totalheight=5.4cm, trim=50 2 80 10, clip=true]{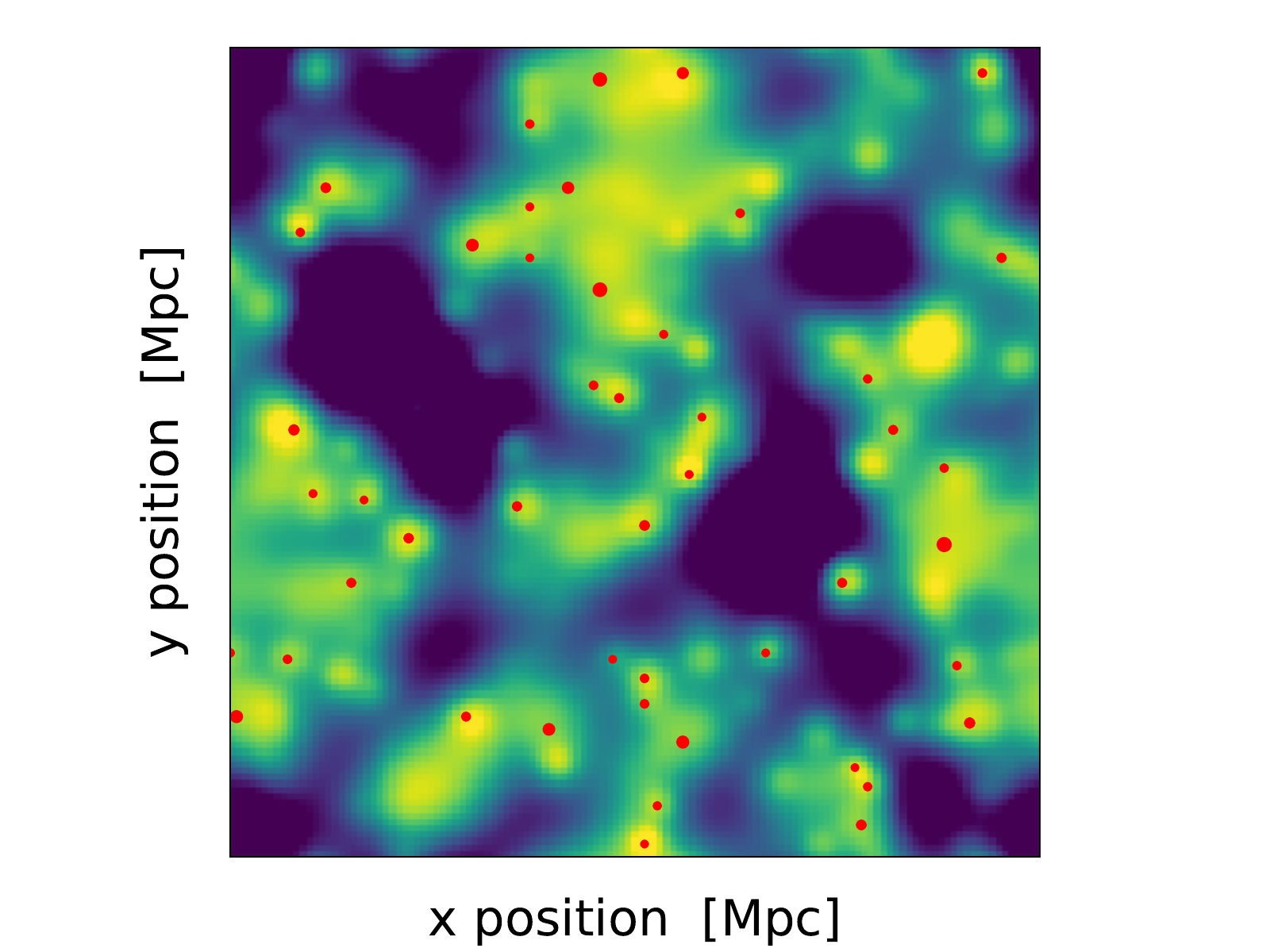}
\includegraphics[totalheight=5.4cm, trim=80 2 80 10, clip=true]{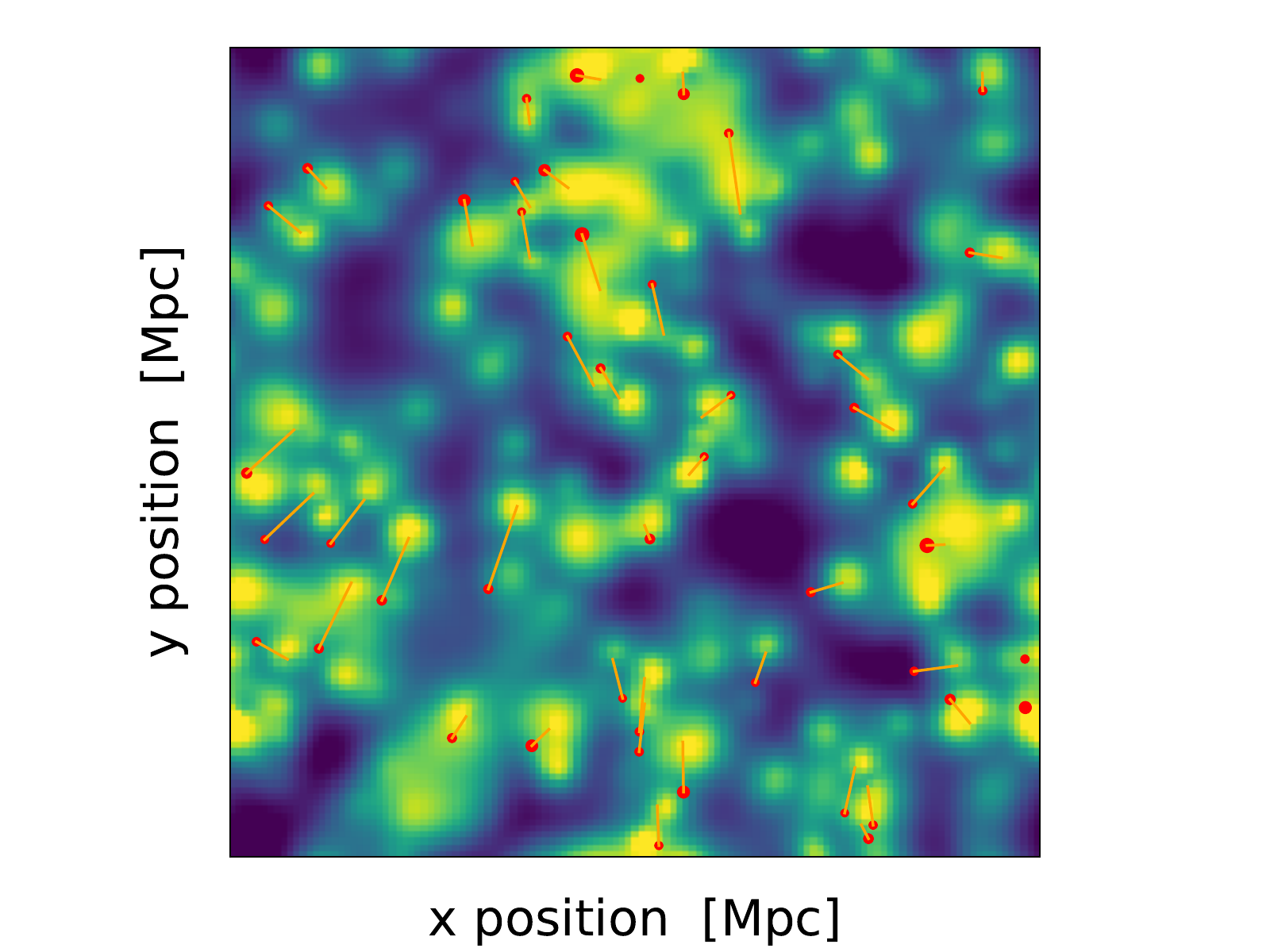}
\includegraphics[totalheight=5.4cm, trim=80 2 80 10, clip=true]{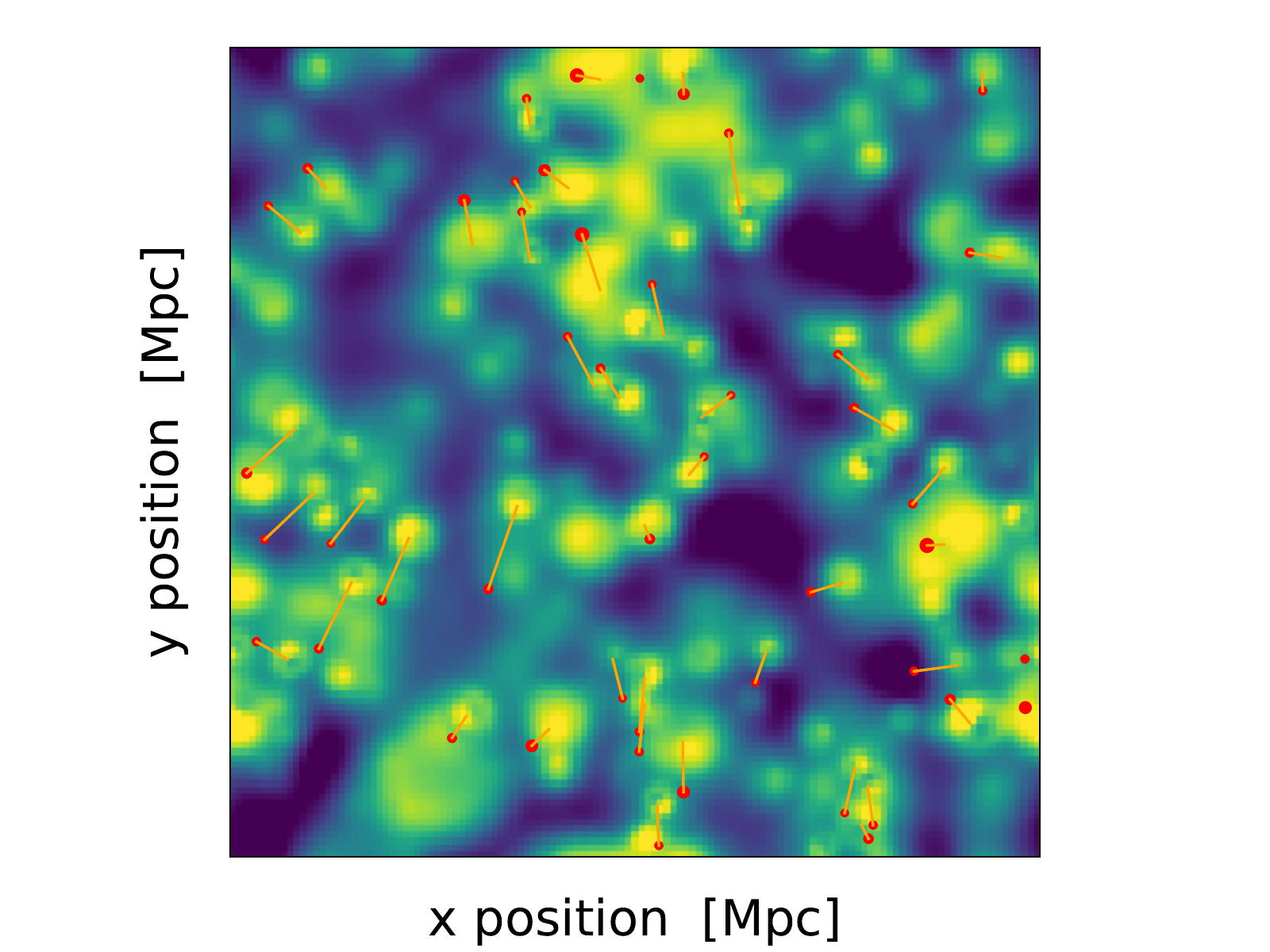}
\end{center}
\caption{The results of our \emph{linear} reconstruction (eqn.~\ref{eqn:linear}) using the Lagrangian position of all $M >10^{12}M_\odot$ halos (L; left panel), this plus their displacements (L+D; middle panel), and this plus the condition that they are density peaks (L+D+P; right panel), for the same linear density field as shown in Figure~\ref{fig:algorithm}.  Each slice is 20 Mpc deep projection through a $100\ $Mpc box, and the image saturates at overdensities of [-3,3].  
 Halo masses and displacements are encapsulated in the same manner as in the rightmost two panels in Fig.~\ref{fig:algorithm}. \label{fig:linearreconstpicture}}
\end{figure}

Linearity means that we can always find a global minimum of the loss function (${L}$), at least once sufficiently conditioned.  
  This section presents solutions to these linear equations for different permutations of the constraint conditions (i.e. of L, D and P in eqn.~\ref{eqn:linear} for $L$), different normalizations of the regularizer, and different galaxy survey specifications.  These equations assume that the positions, the masses of halos, and the Lagrangian displacements $\Psi_{{\rm TRUTH}, j}$ are known perfectly.  Of course, the latter two can only be imperfectly inferred, but this exercise provides a useful bound on reconstruction.   We also compare the resulting reconstructions with those of published nonlinear reconstruction algorithms in \S~\ref{ss:comp} and show that, despite the substantial simplification to reach the linear equations, their cross correlation coefficients with the input linear overdensity field are strikingly similar to those reported in reconstruction studies.  We further show that the linear solutions perform similarly to our best solutions to the nonlinear problem in \S~\ref{sec:nonlinear}.

\begin{figure}
\epsfig{file= 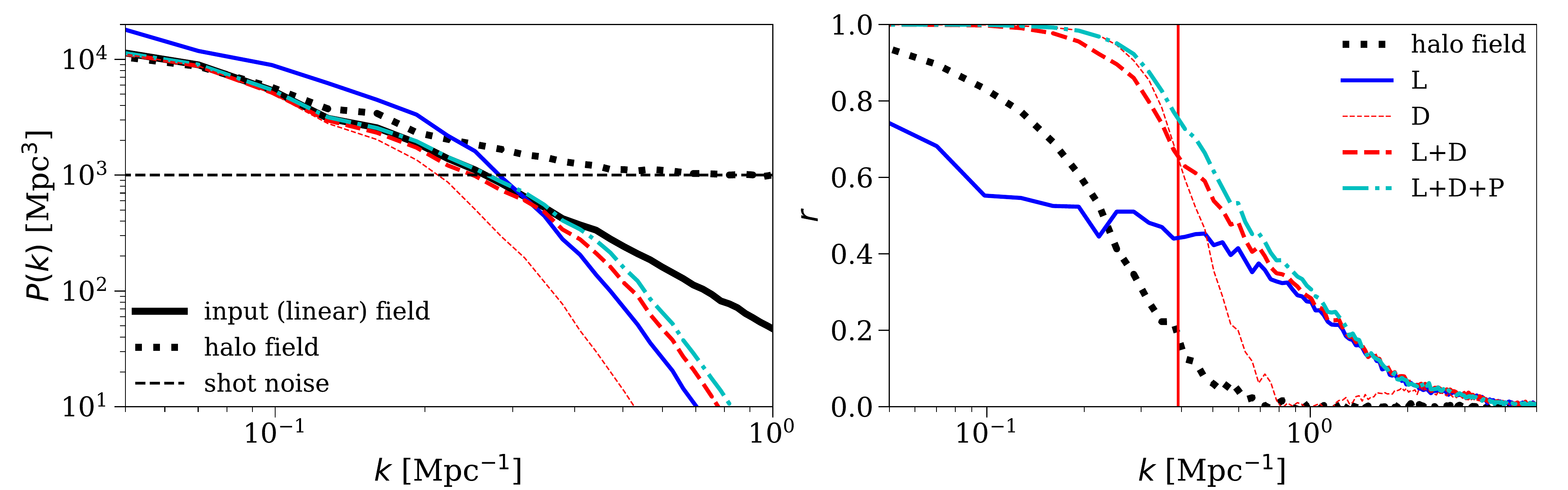, width=15cm}
\caption{The power spectra (left panel) and cross correlation coefficients with the input linear overdensity (right panel), featuring `reconstructions' that use different combinations of the linear constraint conditions.  Curves labeled with `L' use the condition on the Lagrangian-space overdensity threshold (eqn.~\ref{eqn:lagrangian}), `D' use the Zeldovich displacement condition (eqn.~\ref{eqn:displacements}), and `P' use the peaks condition (eqn.~\ref{eqn:peaks}).   These curves are for a reconstruction with $\bar n = 10^{-3}$Mpc$^{-3}$, which is achieved by randomly sampling halos with mass above $10^{12}M_\odot$, which amounts to $45\%$ of all such halos, and they use a Ridge Regression regularization condition with $\sigma=0.01$, although the results are essentially independent of $\sigma$ for $\sigma \lesssim 0.3$.  The horizontal black dashed line in the left panel is the shot noise power calculated as $\bar n^{-1}$, and the red vertical line in the right panel is the estimate from counting displacement constraints, $k_C({\cal N}=3)$.
\label{fig:differentconstraints}
}
\end{figure}

 Figure~\ref{fig:linearreconstpicture} shows images of the reconstructions for different permutations of our linear constraints, conditioning with the Ridge Regression regularizer that is analogous to the Gaussian mode-amplitude prior in other reconstruction studies (eqn.~\ref{eqn:CG}).  (Images of the linear density field that is being reconstructed are shown back in Fig.~\ref{fig:algorithm}.)  The normalization of the regularizer is set to our fiducial value of $\sigma=0.01$, but we will show that our results are insensitive to $\sigma$ as long as $\sigma \lesssim 0.3$.  The lefthand panel in Figure~\ref{fig:differentconstraints} shows the power spectra of the reconstructed input fields from these algorithms (colored curves).  These should be compared with the power spectrum of the true input field (thick black solid curve).  The different colored curves employ different permutations of linear constraints for the reconstruction:  the Lagrangian overdensity condition (L), this condition plus the displacements condition (L+D), both of these conditions plus the peaks condition (L+D+P), and just the displacements condition (D).  Generically, the algorithms reproduce the power spectrum at low $k$ and undershoot at high $k$.  The undershoot at high-$k$ arises because the regularization drives modes that are not significantly constrained to zero, which are generally the high-$k$ ones.  Only in one case shown does the reconstructed power not converge to the input at low wavenumbers, when reconstruction only uses the Lagrangian overdensity condition (L).  In this case, the regularizer preferences smaller mode amplitudes at higher wavenumbers, such that the power at lower wavenumbers must be increased to have sufficient variance on the halo scale to reach the threshold for collapse, $\delta_c(M)$.   Including the constraints on the displacements (L+D) significantly lessens this bias and, indeed, the displacement condition shapes the solution at low wavenumbers (as displacements are the property of the halo field that is the most sensitive in the infrared).  At higher wavenumbers, the Lagrangian overdensity condition shapes the reconstruction, as can be noted by the reconstructed power spectrum using only displacements (D), for which the power dives to zero at lower wavenumbers compared to when the Lagrangian overdensity condition is included.  
  
  Now turn to the righthand panel in Figure~\ref{fig:differentconstraints}, which shows the cross correlation coefficient,
  \beq
  r \equiv \frac{P_{\rm XY}(k)}{\sqrt{P_{\rm XX}(k) P_{\rm YY}(k)}},
  \eeq
where $X$ and $Y$ are the two fields being correlated.  A cross correlation coefficient of unity indicates that the phase of the fields' modes is aligned.   The black thick dashed curve is the cross correlation coefficient between the true halo field and the number density-weighted nonlinear galaxy field (and $r$ is similar in this particular case if we weight by mass).  The other curves in this panel show the cross correlation coefficient between the linear density field and the various reconstructed fields. 
  First, note that all of these reconstructions, with the exception of the Lagrangian overdensity-only (L), find a maximum wavenumber that can be reconstructed with $r\sim 0.5$ that is a factor of $2-3$ larger than where this threshold is met when the galaxy field is simply correlated with the input linear field.\footnote{One could argue that comparing the naive cross correlation is not a fair and one should at least compare a full perturbative bias expansion: however, while adding to the accuracy, the perturbative bias expansion does little to extend the range where $r\approx 1$ \citep{2018arXiv181110640S}.  The gross profile of $r(k)$ between the input overdensity field and nonlinear halo field is set by Poisson noise.}  This factor of $2-3$ improvement is similar to that found in previous reconstruction studies.  However, the Lagrangian overdensity constraint alone is insufficient to reconstruct even the large-scale modes well, falling below $r$ computed from the nonlinear halo field (thick black dashed).\footnote{This comparison is not however a fair one as the Lagrangian overdensity-only (L) reconstruction uses the Lagrangian-space halo field, for which its cross correlation coefficient with the input linear field is well below this reconstructed $r$ on all scales.}  
   Including the displacements immediately makes $r\rightarrow 1$ on large-scales (L+D curve).  Including the peaks constraint (L+D+P) tugs $r$ slightly upward at high wavenumbers, whereas only including the displacement condition (D) results in $r$ transitioning more quickly from one to zero.  
  

\subsection{Comparison of linear reconstruction with the fully nonlinear reconstruction of other studies.}  
\label{ss:comp}
  \begin{figure}
  \begin{center}
\epsfig{file= 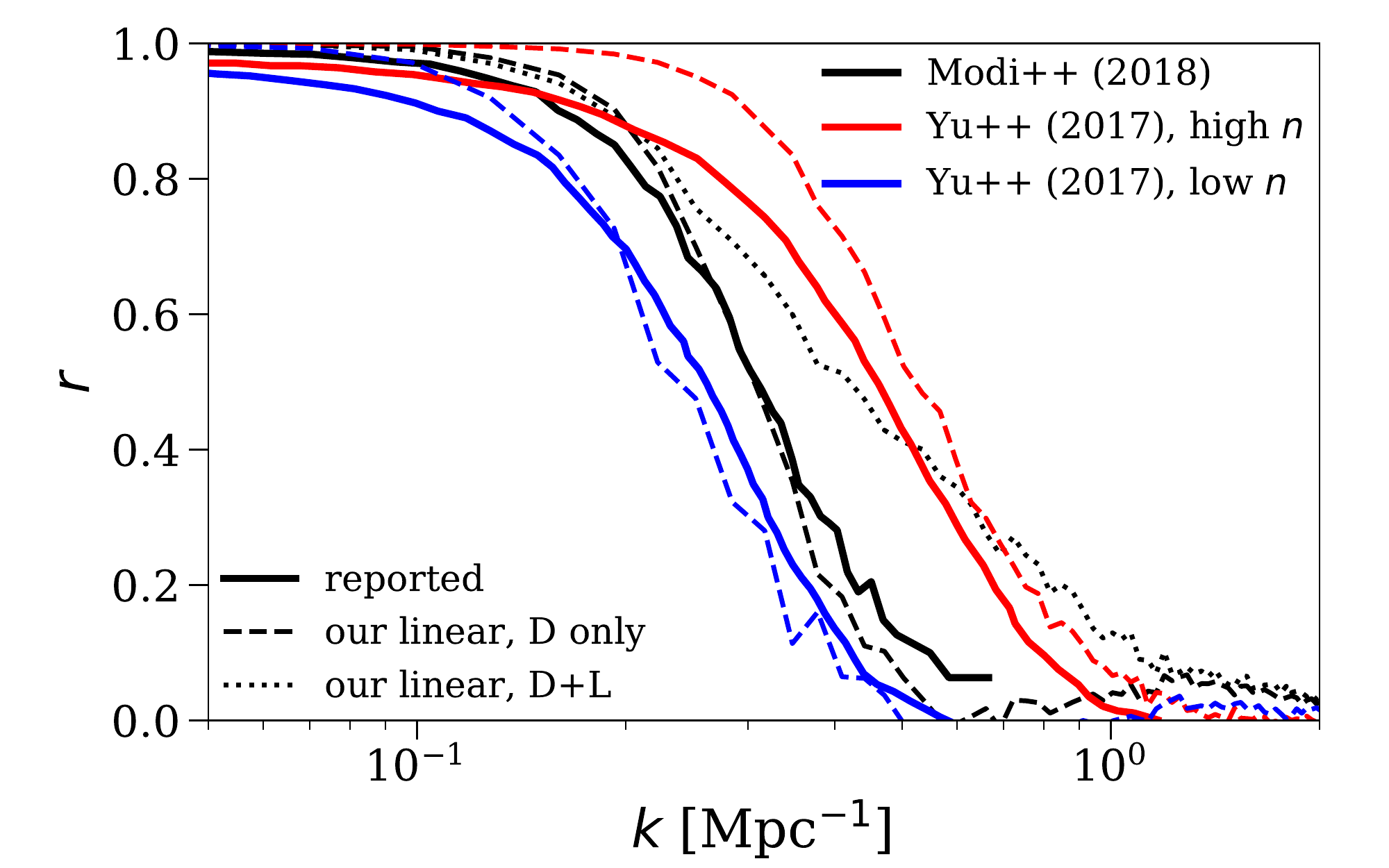, width=10cm}
\end{center}
\caption{{\bf Comparison of our toy reconstruction with the results of prior studies at $z\approx 0.5$.}  The solid curves are the cross correlation coefficients of the input field with the reconstructed presented in Modi et al. \cite{modi18} and in Yu et al. \citep{yu17}.   The dashed curves are the results of the displacement-only linear reconstruction (D) in which we attempt to match the specifications of these studies ($M_{\rm min} =5\times10^{12}M_\odot$, $\bar n=3\times10^{-4}$Mpc$^{-3}$ for Modi et al. 2018 and $M_{\rm min} = 10^{11}M_\odot$, $\bar n=10^{-3}$Mpc$^{-3}$ and $M_{\rm min} = 3\times10^{12}M_\odot$, $\bar n=10^{-4}$Mpc$^{-3}$ for the two samples in Yu et al. 2017).  
  The dotted-curve is the L+D linear reconstruction, which includes the Lagrangian-space overdensity constraint in addition to displacements.  
\label{fig:comparison}
}
\end{figure}

The cross correlation coefficients that our linear reconstruction approximation returns are similar to those found in fully nonlinear reconstruction algorithms.  The solid curves in Figure~\ref{fig:comparison} compare the $r$ of our linear reconstructions at $z=0.5$ with those reported for the reconstructions in Yu et al. \cite[2017,][]{yu17} and Modi et al. \cite[2018,][]{modi18}, which used full cosmological N-body simulations as the input for the nonlinear galaxy field.   The dashed curves show our displacement-only reconstruction (D), attempting to match the halo mass specifications and number densities of the compared studies.  We randomly discard halos above the halo mass threshold to match their number densities.

  Specifically, we compare with the reconstructions in Yu et al. (2017) for their two highest galaxy number densities, which we approximate with the specifications $M_{\rm min} = 10^{11}M_\odot$, $\bar n=10^{-3}$Mpc$^{-3}$ and $M_{\rm min} = 3\times10^{12}M_\odot$, $\bar n=10^{-4}$Mpc$^{-3}$.  They used an `isobaric' reconstruction algorithm that remaps the positions of galaxies to a homogeneous initial field assuming the displacement is curl free.  For both number densities, their $r$ is closest to our displacement-only solution (given by the corresponding dashed curves) as expected.  Yu et al. (2017) considered three redshifts and found little dependence with redshift. 
   A lack of redshift dependence trivially results for our linear reconstruction, as only the halo mass function has redshift dependence in the linear equations and the halo mass function is of little importance for the displacement constraints used for this comparison.  Shell crossing, which erases information in a redshift-dependent manner \cite{2018JCAP...07..043F} and is not captured in our linear model, is one effect that could impart a redshift dependence.  The lack of redshift dependence in Yu et al. suggests shell crossing may not be a principal limitation.
   
Next, let us consider the reconstruction in Modi et al. (2018) for their fiducial specifications of $M_{\rm min} =5\times10^{12}M_\odot$ and $\bar n=3\times10^{-4}$Mpc$^{-3}$.   Modi et al. (2018) used a neural network to extrapolate a low resolution N-body simulation gridded in 4.5~Mpc cells onto a finer grid.  We find that our displacement-only solution agrees well with their result, whereas adding the Lagrangian overdensity constraint results in our $r$ having a tail to higher wavenumbers that outperforms the $r$ in Modi et al. (compare the black solid curve with the black dotted curve).   As their $4.5~$Mpc cells are two times larger than the Lagrangian radius at their minimum halo mass, this lower resolution may make the reconstruction of Modi et al. (2018) less sensitive to the constraints owing to halo peak shapes and, hence, their results better approximated by our displacement-only reconstruction.  Indeed, some calculations that follow suggest that this should be the case.

Both the Yu et al. (2017) and Modi et al. (2018) nonlinear reconstructions do not reproduce the small error found by our linear reconstruction at the lowest wavenumbers.  The low-$k$ reconstruction error of Yu et al.  (Modi et al.) is more or less consistent with being limited by shot noise for number-density (mass-density) weighting.  Our linear reconstructions are able to produce errors well below the shot noise floor at the lowest wavenumbers, a feature we discuss in \S~\ref{ss:regularization}. 

\subsection{(In)sensitivity to regularization parameter $\sigma$ and sub-shot errors} 
\label{ss:regularization}
\begin{figure}
\epsfig{file=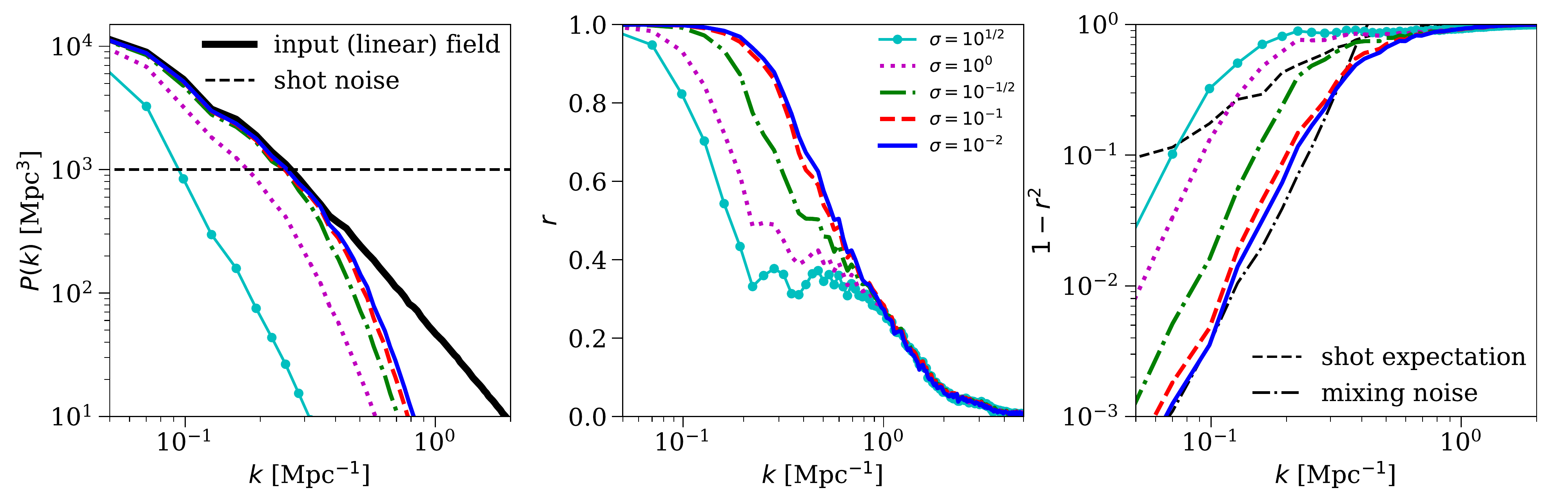, width=15cm}
\caption{{\bf Linear reconstruction for different regularizer normalizations:} The power spectra (left panel), cross correlation coefficients between the input linear overdensity and the reconstructed field $r$ (middle panel), and the $1-r^2$ (right panel).  All reconstructed fields use the linear reconstruction approximation with $N$ Lagrangian overdensity and $3N$ displacement constraints (L+D).   These curves are computed for the case $\bar n = 10^{-3}$Mpc$^{-3}$ by randomly sampling halos with mass above $10^{12}M_\odot$, which requires sampling $45\%$ of all halos in our model.  The dependence of the solutions on our Ridge Regression regularization parameter $\sigma$ only becomes substantial for $\sigma \gtrsim 0.3$.  
  In the noiseless case ($\sigma \rightarrow 0^+$), the reconstruction noise is set by how the low wavenumber modes project onto the set of well constrained eigenvectors, and our estimate for this projection is given by the `mixing noise' (the dot-dashed curve in the rightmost panel).   The mixing noise, which approximates the performance of our linear algorithm, falls substantially below the shot noise expectation, $1-r^2=(\widehat{P}_g \bar n)^{-1}$ where $\widehat{P}_g$ is the galaxy power spectrum in the simulation, shown with the dashed curve in the rightmost panel.
\label{fig:differentsigma}
}
\end{figure}

 All of our reconstruction calculations presented so far adopt $\sigma = 0.01$ for the regularization parameter (c.f.~eqn.~\ref{eqn:linear}).  Larger values of $\sigma$  will increasingly tilt the convex loss function in our linearized problem.  The more poorly a mode is constrained by the galaxy field constraint equations, the more sensitive it should be to the choice of $\sigma$.  However, Figure~\ref{fig:differentsigma} shows that values of $\sigma \sim 1$ are required to have a substantial effect on the cross correlation coefficient. 
   This independence suggests that the modes that are constrained by the galaxy constraints are typically constrained at the level ${\rm var}[\tdelta_{\bfk}] \sim  V P_L(k)$ or better.   
 
 This $\sigma$ independence may seem puzzling, as there are wavenumbers where modes appear to be poorly constrained with $r(k) \ll 1$, yet they are insensitive to even the highest $\sigma$ shown. Indeed, some modes that are reconstructed quite poorly with $r\sim 0.1$ are the least sensitive to $\sigma$!  What reconciles this apparent contradiction is that $\approx 4N$ parameters are well constrained in our reconstructions using the displacement and Lagrangian overdensity constraints (L+D), but these parameters are not necessarily the Fourier modes.  Unconstrained parameters are projected to zero by the regression.  The projection of the Fourier modes onto the constrained eigen-basis shapes the value of the cross correlation coefficient.  

  \begin{figure}
\epsfig{file=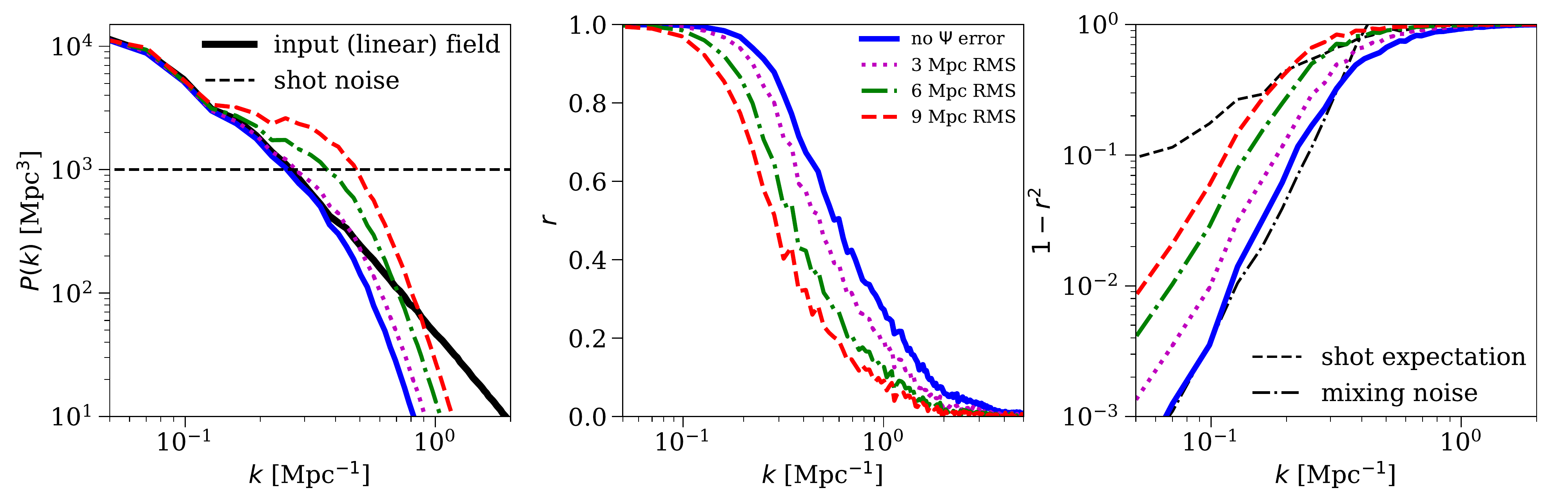, width=15cm}
\caption{{\bf Linear reconstruction when adding noise to $\boldsymbol{\Psi}$:}   The same as Figure~\ref{fig:differentsigma} except, rather than varying the regularization normalization $\sigma$, we randomly add a Gaussian displacement error to the $\Psi_{\rm TRUTH}(\bfq_j)$ in eqn.~\ref{eqn:displacements} before solving the linear equations.  Adding noise to the displacement has a similar effect on $r$ as varying $\sigma$ for reasons described in \S~\ref{ss:regularization}. 
\label{fig:differentRMS}
}
\end{figure}

To understand this mathematically, we can recast our least squares problem in terms of matrix algebra, writing our loss function as
\begin{equation}
L = \sum_{C} ||{\bfA}_C \bftchi - \boldsymbol{b}_C||_2^2 + \sigma^2 || \bftchi ||_2^2, 
\label{eqn:matrixA}
\end{equation}
where the subscript $C$ indexes the constraint conditions (e.g. the Lagrangian overdensity [L], the $j$ component of the displacement [Dj], etc),  $\bftchi \equiv \tdelta_\bfk/\sqrt{V P_L(k)}$, 
 and $|| \ldots ||_2$ denotes the Euclidean norm.  The second term on the right hand side of eqn.~\ref{eqn:matrixA} is our Ridge Regression regularization condition.

For a \emph{single} constraint condition, , which will either be $C={\rm L}$ or $C={\rm D}j$, the solution to eqn.~\ref{eqn:matrixA} is $\widehat{\bftchi}_C = (\bfA_C^\dagger \bfA_C  + \sigma^2 \bfI)^{-1} \bfA_C^\dagger \boldsymbol{b}_C$, where the dagger represents a conjugate transpose. However, if we use singular value decomposition to write $\bfA_C = \bfU \boldsymbol{\Sigma} \bfV^\dagger$, where $\boldsymbol{\Sigma}$ is an $N\times N$ diagonal matrix and $\bfU$ and $\bfV$ are orthonormal matrices, then the solution to eqn.~\ref{eqn:matrixA} with one constraint condition for the modes is
\beq
\widehat{\bftchi} = \bfV \boldsymbol{\Sigma}^\sigma \bfU^\dagger \boldsymbol{b}  ~~~~\text{where}~~~ \Sigma^\sigma_{ij} \equiv \frac{\Sigma_{ii}}{\Sigma_{ii}^2 + \sigma^2} \delta^{\rm K}_{ij},
\label{eqn:regressionSigma}
\eeq
dropping our $C$ subscripts. Thus, Fourier modes that do not project onto the $N$ well-constrained right singular vectors are set to zero.  For modes that do project onto the constrained space, since we find $\widehat{\bftchi}$ does not depend significantly on $\sigma$ for $\sigma < 0.3$, eqn.~\ref{eqn:regressionSigma} indicates that the singular values, $\Sigma_{ii}$, are $\gtrsim 0.3$.   

In the remainder of this section, we use this linear algebra setup to motivate why $\Sigma_{ii} \gtrsim 0.3$ and to derive a formula for the effective noise at low wavenumbers in our linear reconstruction approximation.  Readers interested primarily in the gross properties of the reconstruction can skip ahead to \S~\ref{sec:cosmology}.

\subsubsection{Motivation for singular value sizes}

 We motivate mathematically why the $\Sigma_{ii}$ have such values by first writing out the $\bfA_C$, which we do here for our Lagrangian overdensity and the $j^{\rm th}$ component of the displacement conditions:
\begin{equation}
[A_{\rm L}]_{lm} = \sqrt{\frac{{P_L(k_m)}}{V}} \, e^{-i \bfk_m \cdot \bfq_l}, \label{eqn:AL}
\end{equation}
\begin{equation}
[A_{{\rm D}, j}]_{lm} = -i \ell_{\rm D}^{-1} \sqrt{\frac{{P_L(k_m)}}{  V}} \, \frac{k_{m,j}}{k_m^2} \, e^{-i \bfk_m \cdot \bfq_l},
\end{equation}
where we have ignored factors of ${W}_{M_l}(k)$ as this simplification aids subsequent expressions (and halo exclusion results in a similar effect), and the columns include all wavevectors. The factors of $P_L(k)$ in the above matrices arise from the definition of $\bfA_C$, chosen in order for our mode parameters to be normalized as $\bftchi \equiv \tdelta_\bfk/\sqrt{V P_L(k)}$ such that our regularizer simplifies to the form in eqn.~\ref{eqn:matrixA}.  To understand the size of the singular values $\Sigma_{ii}$, we first calculate
\begin{eqnarray}
\bfA_{\rm L} \bfA_{\rm L}^\dagger =  \begin{pmatrix} \xi_L(R_{{\rm H}, 0}) & \xi_L(\bfq_0 -\bfq_1) &  \xi_L(\bfq_0 -\bfq_2)  & \ldots & \xi_L(\bfq_0 -\bfq_{N-1})  \\           				 \xi_L(\bfq_1 -\bfq_0)  &  \xi_L(R_{{\rm H}, 1})  & \xi_L(\bfq_1 -\bfq_2) & \ldots & \xi_L(\bfq_1 -\bfq_{N-1})\\
                                  \ldots & \ldots & \ldots & \ldots& \ldots \\
                                   \xi_L(\bfq_{N-1} -\bfq_0)  &  \xi_L(\bfq_{N-1} -\bfq_1) & \xi_L(\bfq_{N-1} -\bfq_2) & \dots & \xi_L(R_{{\rm H}, N-1})\\
                 
   \end{pmatrix},
   \label{eqn:ALtAL}
\end{eqnarray}
where $\xi_L(r) \equiv \langle \delta(\bfx) \delta(\bfx + \bfr) \rangle  = V^{-1} \sum_{\forall \bfk} P_L(k) e^{-i \bfk \cdot \bfr}$ is the linear correlation function, and we have approximated the effect of the window functions that we dropped previously by evaluating the diagonal at $R_{{\rm H},j}$, indicating the Lagrangian radius of the $j^{\rm th}$ halo.  The eigenvalues of $\bfA_{\rm L}\bfA_{\rm L}^\dagger $ are equal to the $\Sigma_{ii}^2$, the square of the singular values of $\bfA_L$.  The form of $\bfA_{\rm L} \bfA_{\rm L}^\dagger $ given in eqn.~\ref{eqn:ALtAL} suggests a well conditioned matrix with $\Sigma_{ii}^2 \sim \xi_L(R_{{\rm H}, 0}) \sim 1$:  Two rows in $\bfA_{\rm L} \bfA_{\rm L}^\dagger $ will be most similar when halos are separated by the minimum Lagrangian-space separation our algorithm allows, $\max[R_{{\rm H},i}, R_{{\rm H},j}]$, but we still expect $\xi_L$ to be appreciably different at this minimum separation scale such that the eigenvalue from the eigenvector that principally arises from the subtraction of the two most-similar rows are still likely to be greater than a few tenths.  For modes shaped by the Lagrangian overdensity constraint, this explains why $\sigma \sim 1$ demarcates where the normalization of the regularization condition starts to substantially affect the solution.     

Now, if we repeat and compute $ \bfA_{{\rm D}, j} \bfA_{{\rm D}, j}^\dagger$, the same expression as eqn.~(\ref{eqn:ALtAL}) holds but with $\xi_L \rightarrow \ell^{-2}_D \langle \psi_j(\bfx) \psi_j(\bfx + \bfr) \rangle = \ell^{-2}_D V^{-1} \sum_{\forall \bfk} k_j^2 k^{-4} P_L(k) e^{-i \bfk \cdot \bfr}$, i.e the correlation function of displacements along the $j$ direction measured in units of $\ell_{\rm D} = 10~$Mpc.  We note that the average displacement is of the order of $10~$Mpc such that again the diagonal values of $ \bfA_{{\rm D}, j} \bfA_{{\rm D}, j}^\dagger$ are approximately unity.  
  The matrix $ \bfA_{{\rm D}, j} \bfA_{{\rm D}, j}^\dagger$ is not as well conditioned as $\bfA_{\rm L} \bfA_{\rm L}^\dagger$ because displacements are more correlated between galaxy positions.  The dependence of our reconstructions on $\sigma$ shown in Fig.~\ref{fig:differentsigma} suggests that the smallest singular values of $\Sigma_{ii}\approx 0.1$, but with many values around $\Sigma_{ii}\sim 1$ and with the lowest wavenumbers projecting onto the singular vectors that have the largest $\Sigma_{ii}$.


  
 Thus, reconstruction measures nearly as many numbers as constraints once $\sigma^2 \lesssim 0.1$.  This statement can be related to errors on, for example, the displacements -- when they are constrained to a few Mpc (as our singular values in this case are in units of $10~$Mpc), $3N$ numbers are measured.  Indeed, Figure~\ref{fig:differentRMS} examines the reconstructed field that results when adding an uncorrelated Gaussian error to each halo's displacement, an action which has a similar effect on $r$ as varying $\sigma$.   However, the $XN$ well-constrained right-singular vectors are not necessarily the Fourier modes we desire to measure.  
   Indeed, the right singular vectors for the Lagrangian overdensity problem do not appear to project well onto Fourier modes as evidenced by the $r$ values being substantially off unity at all wavenumbers (see the blue solid curve in the righthand panel of Fig.~\ref{fig:differentconstraints}).  When displacements are included, the well constrained subspace appears to be better represented by the lowest wavenumber Fourier modes.  Interestingly, the projection is such that the lowest wavenumbers are substantially better constrained than the shot noise expectation (see the right panel in Fig.~\ref{fig:differentsigma}).  We aim to understand this projection better in what follows. 

\subsubsection{Small wavenumber limit of linear reconstruction}
\label{sec:smallkprojection}

 For a better handle on this projection, rather than considering $\bfA_{C} \bfA_{C}^\dagger$ as above, we instead compute $\bfA_{C}^\dagger  \bfA_{C}$.  The eigenvectors of this matrix are the same as the right singular vectors of $\bfA_{C}$.  Since the displacements shape the reconstruction at low wavenumbers, let us consider the $j^{\rm th}$ component of the displacements, where
%
\begin{equation}
\left [\bfA_{Dj}^\dagger  \bfA_{Dj}\right]_{sm} =  \ell_{\rm D}^{-2}  \sqrt{\frac{P_L(k_s )P_L(k_m )}{V^2} }  \frac{k_{s,j} k_{m,j}}{k_s^2k_m^2}  \left[N \delta^{\rm K}_{sm} + \left(\bar n b_{Lq}  \tilde  \delta_{(\bfk_s - \bfk_m)}  + G(\sqrt{N}) \right)(1-\delta^{\rm K}_{sm}) \right].
\end{equation}
We have used that $\sum_{j=0}^N e^{-i (\bfk_s - \bfk_m) \cdot \bfq_j}=  \bar n b_{Lq}  \tilde  \delta(\bfk_s - \bfk_m ) + G(\sqrt{N})$ for $\bfk_s\neq \bfk_m$, where $b_{Lq}$ is the galaxy Lagrangian bias and $G(\sqrt{N})$ is a `random' c-number with standard deviation $\sqrt{N}$ that owes to shot noise (i.e. you can think of the Riemann sum as a Monte-Carlo integral of a continuous overdensity field and $G(\sqrt{N})$ is the error).

We can estimate how much of a rotation of the basis is required to diagonalize $\bfA_{Dj}^\dagger  \bfA_{Dj}$, where the vectors that describe these rotations are the eigenvectors.  If we only consider high wavenumbers where typically  $|G(\sqrt{N})| > |\bar n b_{Lq} \tilde \delta | $ such that shot noise dominates, in the plane indexed by $(\bfk_s, \bfk_m)$ a rotation by an angle of 
\begin{equation}
\theta_{\bfk_s \bfk_m} \approx \sqrt{\frac{P_L(k_m )}{P_L(k_s)}} \frac{k_s}{k_m} \frac{G(\sqrt{N})}{N},
\label{eqn:shottheta}
\end{equation}
is required to diagonalize from the Fourier basis. Let us assume that any rotation that mixes a mode $\bfk_s$ with modes with $k_m>k_C$ results in that component of the mode being lost, motivated by our numerical displacement-only reconstructions in which modes with $k \lesssim k_C$ are well constrained whereas modes with $k \gtrsim k_C$ are poorly constrained (e.g. Fig.~\ref{fig:differentconstraints}).   Under this assumption and further assuming that shot noise dominates these modes (justified in concordance cosmology because $k_S \approx k_C$; \S~\ref{sec:motivations}), we can use eqn.~(\ref{eqn:shottheta}) to estimate the cross correlation coefficient
\begin{equation}
r(k_s) = \prod_{\forall |\bfk_m|>k_C}   \cos[\theta_{\bfk_s \bfk_m}] = \left(1 - \frac{1}{2}\sum_{\forall |\bfk_m|>k_C} \theta_{\bfk_s \bfk_m}^2  + \ldots \right),
\end{equation}
 or
 \begin{equation} 
 1 -r(k_s)^2 \approx \sum_{\forall |\bfk_m|>k_C} \theta_{\bfk_s \bfk_m}^2 \approx \frac{k_s^2}{\bar n P_L(k_s)} \int_{k_C}^{\infty} \frac{d^3k}{(2\pi)^3}\frac{P_L(k)}{k^2} ,\label{eqn:sens1}
\end{equation} 
using that modes are pixelated with $(\Delta k)^3 = (2\pi)^3/V$.    If we roughly expect $ 1 -r^2$ from the three displacement conditions to be reduced by a factor of three relative to the single component estimate given by eqn.~\ref{eqn:sens1}, which results in the low-wavenumber estimate of
 \begin{equation} 
  1 -r(k)^2 \approx  2000 {\rm Mpc^5} \times \frac{k^2}{P_L(k)} \frac{ \sigma_{\Psi}(k_C)^2}{2 {\rm ~Mpc}^2} \left(\frac{10^{-3} {\rm Mpc}^{-3}}{\bar n}\right)  \text{~~~[\bf Mixing Noise Limit]},
  \label{eqn:sensitivity}
 \end{equation} 
 where $ \sigma_{\Psi}(k_C)^2$ is defined by eqn.~\ref{eqn:sigmadisp}, and $ \sigma_{\Psi}(k_C)^2 = 2~$Mpc$^{2}$ for $k_C = 0.5$Mpc$^{-1}$ in $\Lambda$CDM.  The rightmost panel in Figure~\ref{fig:differentsigma} shows that this expression for the reconstruction noise, which we term `mixing noise', captures the $r$ we find at low $k$ in the low $\sigma$ limit in which the solution is not affected by the normalization of the regularizer.  Additionally, we will show in \S~\ref{sec:cosmology} that the mixing noise limit formula applies even in cosmologies where $P_L$ is very different.
 


Finally, while the limiting sensitivity given by eqn.~\ref{eqn:sensitivity} becomes independent of the regularization condition normalization, $\sigma$, once $\sigma \lesssim 0.3$, our regularization condition does shape our results as it determines the weighting via $\widetilde{\chi}_j = \tdelta_{\bfk_j}/\sqrt{VP_L}$.  This weighting is responsible for the factors of $P_L$ in eqn.~\ref{eqn:sens1}.  If we more generally use $\widetilde{\chi}_j = \tdelta_{\bfk_j}/\sqrt{V{\cal W}(k)}$ then eqn.~\ref{eqn:sens1} would instead become
 \begin{equation} 
 1 -r(k_i)^2 \approx  \bar n^{-1} \frac{k_i^2 }{{\cal W}(k_i)} \int_{k_C}^{\infty} \frac{d^3k}{(2\pi)^3}\frac{{\cal W}(k_j)}{k_j^2},
 \label{eqn:rW}
\end{equation} 
with the caveat that our approximations become less good as ${\cal W}(k)$ is made to be more strongly increasing with $k$ than $\sqrt{V P_L}$.  While ${\cal W}(k) = P_L(k)$ may be the most natural weighting, eqn.~\ref{eqn:rW} suggests that one could design ${\cal W}(k)$ to result in even much smaller error bars by increasing its tilt towards the infrared.  The reason why there is no free lunch is that the matrix $ \bfA_{{\rm D}, j} \bfA_{{\rm D}, j}^\dagger$ becomes more poorly conditioned the more red-sensitive that ${\cal W}(k)$ becomes, amplifying modeling errors.  

%

  \subsection{Altering the cosmology} 
 \label{sec:cosmology}
 
  \begin{figure}
  \begin{center}
\epsfig{file= 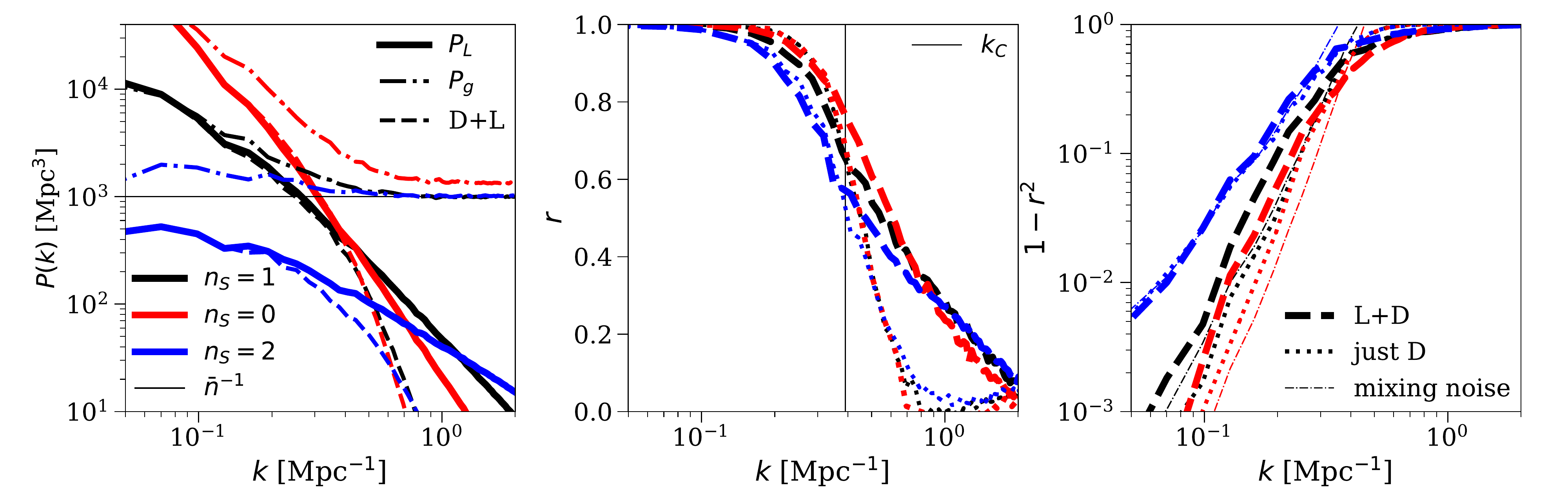, width=16cm}
\end{center}
\caption{{\bf Reconstruction varying cosmology ($n_s$) in the linear approximation:} The power spectra (left panel), the cross correlation coefficients between the input linear overdensity and reconstructed fields $r$ (middle panel), and $1-r^2$ (right panel), varying the effective tilt of the input power spectrum by $\pm 1$ from the fiducial $n_s\approx1$ cosmology in a manner such that the overdensity variance is still the same in a tophat window with mass of $10^{11} M_\odot$ (which roughly matches $n(>M)$ and $b_L$  for the three cosmologies).  All reconstructed fields use our linear reconstruction approximation with the Lagrangian overdensity plus displacement conditions (L+D; dashed curves) or just displacements (D; dotted curves).    In addition to the reconstruction results, the left panel shows the input power spectra (solid curves) and the galaxy power spectra (dot-dashed curves), the middle panel shows the constraints scale ($k_C$) at which $3N$ modes have smaller wavenumbers, and the right panel shows the mixing noise (eqn.~\ref{eqn:sensitivity}) for the three cosmologies.   These curves are computed for the case $\bar n = 10^{-3}$Mpc$^{-3}$ by randomly sampling halos with mass above $10^{12}M_\odot$ (which requires sampling $100\%$ of halos in the reddest tilt, half of them in the fiducial, and a quarter of them in the bluest).  
\label{fig:cosmology}
}
\end{figure}
  
  Figure~\ref{fig:cosmology} shows how the linear reconstruction fares for different spectral tilts of the primordial matter overdensity power spectrum, changing the effective spectral index by $\pm 1$, but in a manner that the density variance in a tophat sphere with mass $10^{11} M_\odot$ is the same (such that the $n(>M)$ and $b_L$ are approximately matched for galactic halos for these three cosmologies).  We compare these cosmologies with $\bar n = 10^{-3}$Mpc$^{-3}$ by randomly sampling halos with mass above $10^{12}M_\odot$, which requires sampling $100\%$ of halos in the reddest tilt, half in the fiducial, and a quarter in the bluest tilt cosmology.
  
 The left panel shows the power spectra of the input linear overdensity (solid curves), of the reconstruction in the linear approximation (dashed curves), and of the galaxy field (dot-dashed curve) for the three different cosmologies.  The fractional contribution of shot noise is much different in these exotic `tilted' cosmologies, with the one with the bluest tilt boasting a galaxy power spectrum that is significantly shaped by shot noise at all simulated scales.  All cosmologies however have the same constraints scale $k_C$, and $r=0.5$ is achieved at nearly the same wavenumber in all three linearized reconstruction calculations (middle panel; the dashed curves show L+D and the dotted curves show D).  Even the approximate profile of the cross correlation coefficient is similar between the different calculations.  Clearly shot noise is not what sets its profile.  Rather, the quick transition from $r\approx 1$ to $r\approx 0$ is primarily set by mode counting, where the wavenumber that has $3N$ shorter modes, $k_C$, is shown as the vertical line in the middle panel.
 
 The rightmost panel in Figure~\ref{fig:cosmology} now shows $1-r^2$, where the dashed and dotted curves are respectively the L+D and D linearized reconstructions.  Reconstruction in the linear approximation fares far better than the shot noise `floor', with a significantly steeper wavenumber scaling in the three cosmologies than if the noise were set by shot noise.  This is particularly evident in the bluest tilted cosmology, where shot noise would result in an error that is nearly wavenumber independent.  The blue thin dot-dashed curve show our `mixing noise' estimate in this shot noise-dominated cosmology (eqn.~\ref{eqn:sensitivity}), which remarkably reproduces the salient behavior in all three cosmologies.\footnote{We cap our integral to calculate $\sigma_\psi^2$ at $k_M$ for $10^{12}M_\odot$ or otherwise the curve would shift up by almost a factor of two for the $n_s=2$ case.}

\subsection{Number density and halo mass dependences}
\label{ss:dependences}
 \begin{figure}
\epsfig{file= 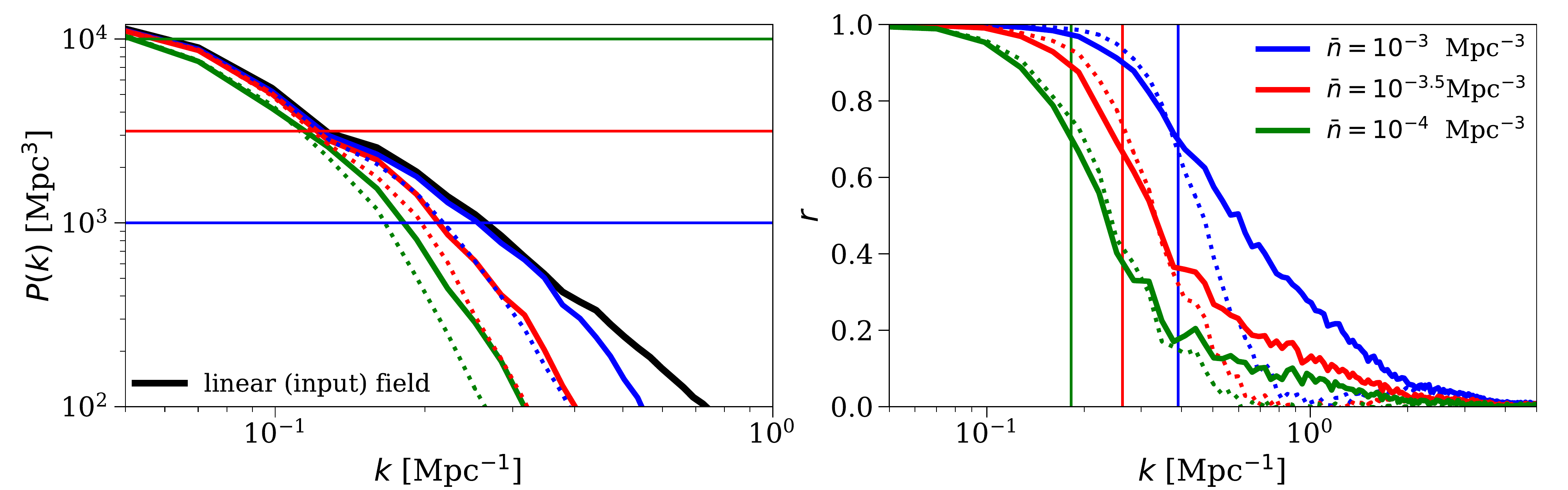, width=15cm}
\caption{{\bf Reconstruction varying number density in the linear approximation:}  The power spectra (left panel) and cross correlation coefficients with the input (right panel) for three halo number densities in our linear reconstruction approximation.  Number densities are selected by randomly sampling from the halos that have masses greater than $10^{12}M_\odot$.  The colored solid curves are linear models that include both the Lagrangian overdensity and displacement constraints (L+D), and the dotted curves with the corresponding color only use the displacements (D).  The horizontal lines in the left panel are the $\bar n^{-1}$, and the vertical lines in the right panel show the constraints wavenumbers, $k_C$.
\label{fig:differentnumbers}
}
\end{figure}

Figure~\ref{fig:differentnumbers} shows how the reconstructed power and cross correlation coefficients vary with the galaxy number density, $\bar n$, using our linearized algorithm.  Number densities are fixed by randomly sampling a fraction of halos that have masses greater than $10^{12}M_\odot$, maintaining the same spectrum of halo masses.   A factor of ten increase in $\bar n$ results in a factor of $10^{1/3} \approx 2$ improvement in the wavenumber reach, and there is the analogous trend for a factor of ten smaller $\bar n$.  The solid vertical lines in this figure are the constraints scale $k_C=(6\pi^2 \bar n)^{1/3}$.  The low $k$ behavior of $1-r^2$ scales inversely with number density and is well explained by our mixing noise estimate (eqn.~\ref{eqn:sensitivity}).

  \begin{figure}
\epsfig{file= 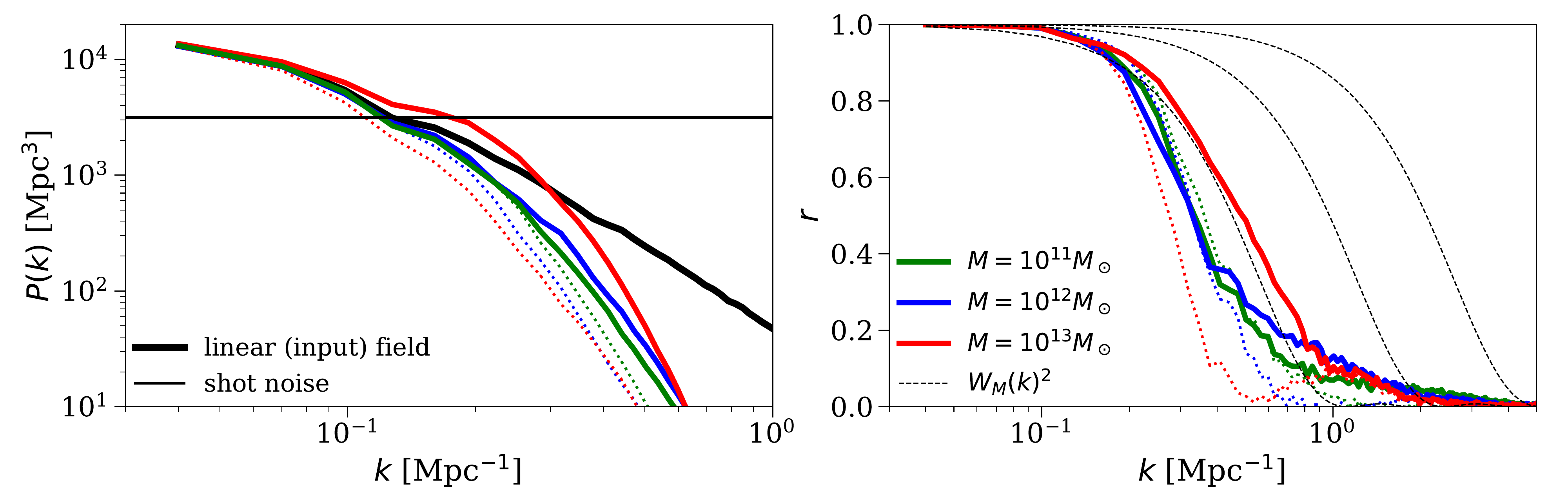, width=15cm}
\caption{{\bf Reconstruction varying halo masses in the linear approximation:}   The power spectra (left panel) and cross correlation coefficients with the input (right panel) for three minimum halo masses and a fixed number density of $\bar n = 3\times10^{-4}$Mpc$^{-3}$ in our linear reconstruction approximation.  This number density is fixed by randomly choosing a fraction of halos above the halo mass threshold. The solid curves are linear models that include both the Lagrangian overdensity and displacement constraints (L+D), and the dotted curve use only the displacements (D), with the color indicating the halo mass threshold.  The thin black dashed curves in the right panel show ${W}_{M}(k)^2$, the square of the real-space tophat window functions, for $M=10^{13}, ~10^{12},$ and $10^{11}M_\odot$ (from left to right respectively).  
\label{fig:differentmasses}
}
\end{figure}

Figure~\ref{fig:differentmasses} fixes the number density to $3\times10^{-4}$Mpc$^{-3}$ and varies the minimum halo mass, considering minimum masses of $M=10^{11}, ~10^{12},$ and $10^{13}M_\odot$.  Also shown is the square of the square of the real-space tophat window functions that enclose a mass, $W_{M}(k)^2$, for these three $M$ (black dashed curves).  Generally the wavenumbers of modes that are constrained are smaller than the halo scale where $W_{M}(k)^2$ transitions to zero.  In these cases, the displacement-only reconstruction is similar to adding the Lagrangian overdensity constraints aside from a high-$k$ tail of low $r$ values.  When the number densities are sufficiently high such that $k_C$ becomes more comparable to the halo scale, the Lagrangian overdensity constraints become more effective at boosting the span with $r>0.5$ (as well as resulting in an overshoot in the reconstructed power), as can be seen for the $10^{13}M_\odot$ case.  We suspect there is a similar behavior when the reconstruction algorithm does not resolve the Lagrangian halo scale $R_{{\rm H}}$ -- specifically that the resulting $r$ mimics the displacement-only reconstruction.   This suspicion could explain why the displacement-only reconstruction best matches the Modi et al. (2018) \citep{modi18} reconstruction, as their algorithm does not resolve $R_{{\rm H}}$.

We have also investigated adding a lognormal scatter to the mass associated with halo (which enters in the Lagrangian  peak condition and not the displacement condition) as would arise from uncertainty in the inference of halo masses.  We find that this scatter has the effect of reducing the benefits of adding the Lagrangian overdensity condition.  This scatter has little effect at low wavenumbers, as the displacements drive the reconstruction there.  Reconstruction algorithms that use a gridded halo field may be more sensitive to such halo mass uncertainties (see \S~\ref{sec:nonlinearmodel} ).

\section{Reconstruction in the nonlinear model}
\label{sec:nonlinear}\label{sec:nonlinearmodel}

So far we have concentrated on understanding reconstruction in our convex, linear limit.  This section finally investigates the solution to the nonlinear problem!  The main result is that, with our best nonlinear algorithms, we find a cross correlation coefficient between the input field and the reconstructed field that has a similar profile to that found with our linear approximation.  Our nonlinear solutions do make larger errors at low wavenumbers that are most notable when we consider $1-r^2$, but improvements may be possible there.

 The loss function for our full nonlinear toy problem is
\begin{eqnarray}
{L} \equiv  {\sum_{j=0}^{N}  \left(V^{-1} \sum_{\forall  \bfk} \tdelta_{\bfk} e^{-i \bfk \cdot \left[ \bfx_j - \boldsymbol{\psi}(\bfq_j |\tdelta_{\bfk'})  \right]} W_{M_j}(k) - \delta_c(M_j) \right)^2} +\sigma^2 \sum_{\forall \bfk}  \frac{|\delta_\bfk|^2}{P_L(k) V}.
\end{eqnarray}
The first term is just repeating our model (eqn.~\ref{eqn:master}) and the second is the same regularization as before.\footnote{We have investigated including the condition that halos form at peaks at their Lagrangian position, but find no improvement (and rather somewhat poorer performance) and so do not include this condition in our nonlinear model. }  We again highlight the simplification we use for our nonlinear problem where we evaluate the estimated displacement in the exponent at the true Lagrangian position $\bfq_j= \bfx_j -\boldsymbol{\psi}_{\rm TRUTH}$ rather than estimating $\bfq_j$ from the estimated displacement field $\boldsymbol{\psi}$ and the final position of the galaxy $\bfx_j$ via some optimization method, a primitive one being Taylor expansion.  We suspect that doing the optimization so that our result has no dependence on the unobservable quantity, $\bfq_j$, would not affect the solution relative to this simplification (see \S~\ref{sec:toyuniverse}).  Beyond this simplification, the inputs for solving the nonlinear model are the `observables', $\bfx_j$ and $M_j$ for  $j \in [0,N)$.  As in previous sections, we apply this formalism to our toy realization of the universe described in \S~\ref{ss:excursionset}.

 This nonlinear loss function allows us to comment on what $\tdelta_{\bfk}$ are likely to fall near the convex subspace that encompasses the true solution.  The wavenumbers $k$ that contribute to $\sum_{\forall  \bfk} \tdelta_{\bfk} e^{-i \bfk \cdot \bfq_j} W_{M_j}(k)$  are generally larger than those that contribute to $\boldsymbol{\psi}(\bfq_j |\tdelta_{\bfk'})$ and smaller than $\sim R_{{\rm H}}^{-1}$, where $R_{\rm H}$ is the Lagrangian size of  the halo.  Once $\boldsymbol{\psi}(\bfq_j |\tdelta_{\bfk})$ is estimated with accuracy $R_{{\rm H}}$, the exponential factor $e^{-i \bfk \cdot \bfq_j}$ is no longer very sensitive to the mode amplitudes  (allowing one to expand the part of the argument that still depends on $\boldsymbol{\psi}$) and, to the extent $\boldsymbol{\psi}(\bfq_j |\tdelta_{\bfk'})$ arises from smaller wavenumber modes than $\sum_{\forall  \bfk} \tdelta_{\bfk} e^{-i \bfk \cdot \bfq_j} W_{M_j}(k)$, the problem becomes convex.  In the relevant limit where lower wavenumbers are reconstructed best, the wavenumbers that need to be accurately reconstructed to estimate  $\boldsymbol{\psi}(\bfq_j |\tdelta_{\bfk})$ to within $R_{{\rm H}}$ are $k<k_{\rm disp}$ (\S~\ref{sec:motivations}).  Fig.~\ref{fig:scales} shows that $k_{\rm disp}\approx 0.3\;$Mpc$^{-1}$ for a survey consisting of all halos with $M>5\times10^{12}M_\odot$.  This $k_{\rm disp}$ is not much larger than the wavenumbers at which the galaxy field correlates well with the input field and, thus, a reconstruction algorithm can be initialized with these modes close to their correct values.  This argument helps explain the seemingly convex behavior of reconstruction in the literature in which even gradient descent-like algorithms do not seem to get stuck in minima that are far from the input field.\footnote{Of course, for modes well described by linear theory, one expects the reconstruction problem should be trivially convex in the region considered for the $\tdelta_{\bfk}$; we are referring to the more nonlinear modes.}  However, if the halo overdensities owe to modes that are entirely distinct from those that contribute to the displacements, reconstruction could `draw' a halo anywhere and, therefore, would not be able to constrain the halo displacements.  Thus, the intermediate modes that contribute to both $\boldsymbol{\psi}$ and $\boldsymbol{\psi}(\bfq_j |\tdelta_{\bfk'})$ must play a key role.

To illustrate these points, we now present reconstructions in our simplified nonlinear problem.  Because the loss function is not likely to be convex far from the true solution, the reconstruction is likely to be more successful if it starts from a point that is as good a guess as possible for the evolved field.  We consider three starting points:
\begin{description}
\item[zero:] $\tilde{\delta}(\bfk) = 0$.  For this distant starting point, the L-BFGS algorithm gets trapped relatively far away from the minimum found by the linear algorithm presented in the previous section. 
\item[realistic:] The strategy for this case is to guess a field that is somewhat close to the input overdensity field using quantities that are observationally accessible.  Starting with an overdensity field with 
\begin{equation}
\tdelta^{(0)}(\bfk) = \tilde{\delta}_{\rm g}^{\rm mw}(\bfk)/b_g ~~~~~~\text{for $k < 0.2~$Mpc$^{-1}$ } \nonumber
\end{equation}
and $\tdelta^{(0)}(\bfk) =0$ otherwise. Here $\tdelta_{g}^{\rm mw}$ is the mass-weighted galaxy field.\footnote{The galaxy bias $b_g$ is estimated as $\widehat{b}_g= {\cal N}_k^{-1} \sum_{|k|<k_{\rm max}} \tdelta_{g}^{\rm mw} \tdelta_{\rm TRUTH}^*/(\tdelta_{\rm TRUTH} \tdelta_{\rm TRUTH}^*)$, where $ \tdelta_{\rm TRUTH}$ is the input density field.  This bias estimate uses the ${\cal N}_k$ lowest modes in the simulation that satisfy $k< k_{\rm max} = 4\pi/V^{1/3}$.  Using the input to calculate the bias is not something an observation would have access to, but a realistic survey would have a much larger volume than our $200~$Mpc box to measure the bias.}     Next, we take the halo field and from the position of halos displace backwards to get the Lagrangian position, with the displacements calculated from $\tdelta^{(0)}(\bfk)$, and place a compensated overdensity profile at $\bfq_j^{(0)} = \bfx_j - \boldsymbol{\psi}(\bfq_j | \tdelta^{(0)}_\bfk)$ given by
\begin{equation}
{\cal W}^c_{M_j}(\bfq, \bfq_j,  \delta(\bfq_j))=   2 \left[\delta_c(M_j) - \delta(\bfq_j^{(0)})\right] \exp\left [ - \frac{(\bfq - \bfq_j^{(0)})^2}{2 R_{{\rm H}}(M_j)}\right] \left(1 - \sqrt{\frac{\pi}{2}}\frac{|\bfq - \bfq_j^{(0)}|}{2 R_{{\rm H}}(M_j)} \right),
\end{equation}
where $\delta_c(M_j)$ is the overdensity that defines our halo (eqn.~\ref{eqn:master}) and $\delta(\bfq)$ is the matter overdensity field. The spatial integral over ${\cal W}^c_{M_j}$ is zero,\footnote{In practice, our ${\cal W}^c_{M_j}$ only goes out $3.5 R_{{\rm H}}(M_j)$ from $\bfq_j$ as this contains 90\% of the support.} and this fact also means that these profiles do not alter the power at low enough $k$.  
  Thus, the matter overdensity input field for the L-BFGS algorithm is given by $\delta^{(N)}$ where
\begin{equation}
\delta^{(n)}(\bfq) = \delta^{(0)}(\bfq) + \sum_{j=0}^n {\cal W}^c_{M_j}(\bfq, \bfq_j^{(0)},  \delta^{(j)}(\bfq)),
\end{equation}
and we organize the placement such that the $M_j$ that are summed are in decreasing order.
For $M=5\times10^{12}M_\odot$ ($1\times10^{12}M_\odot$), the value of $L$ for this input $\tdelta_\bfk$ is reduced by a factor of 3(2) over the {\bf zero} initialization with $\tdelta_\bfk =0$.  
\item[idealized:] We use the identical scheme as (2) except with $\tdelta^{(0)} = \tdelta_{\rm TRUTH}$ for $k < 0.2~$Mpc$^{-1}$ and $\tdelta^{(0)}(\bfk) =0$ otherwise (i.e. we start with the correct solution at low wavenumbers).  This idealized starting point results in a similar initial value for $L$ to the \emph{realistic} algorithm.
\end{description}
All of the results presented in this section use $\sigma=0.01$; we find a similar dependence on $\sigma$ as in our linear reconstructions.

 \begin{figure}
\begin{center}
{\includegraphics[width=13.4cm]{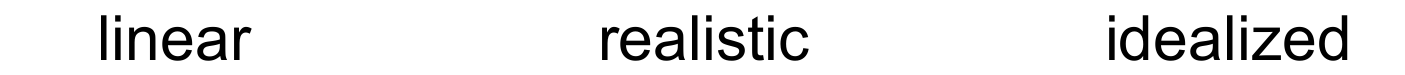}}
\includegraphics[totalheight=5.4cm, trim=50 2 80 10, clip=true]{lindisp.pdf}
\includegraphics[totalheight=5.4cm, trim=80 2 80 10, clip=true]{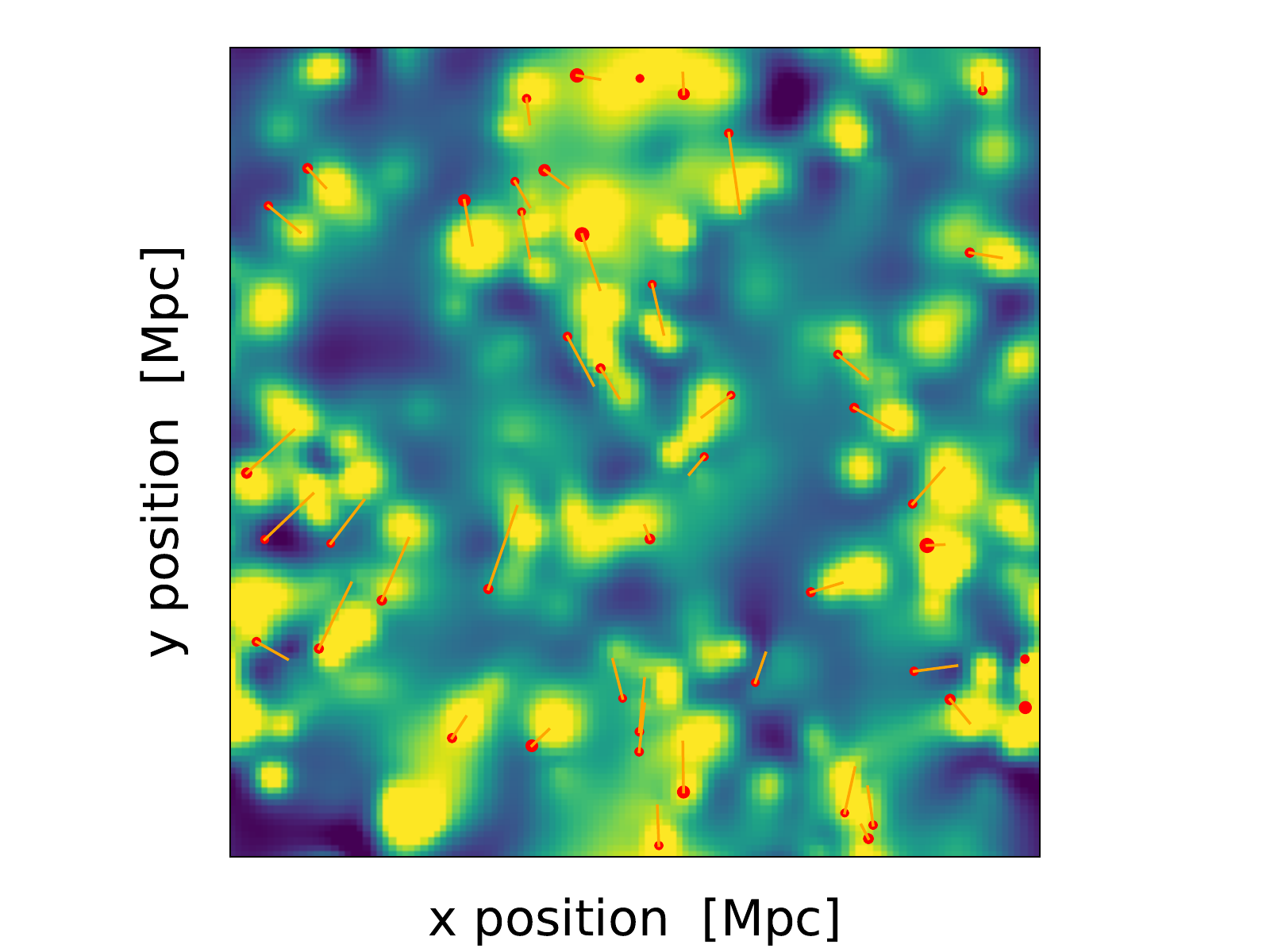}
\includegraphics[totalheight=5.4cm, trim=80 2 80 10, clip=true]{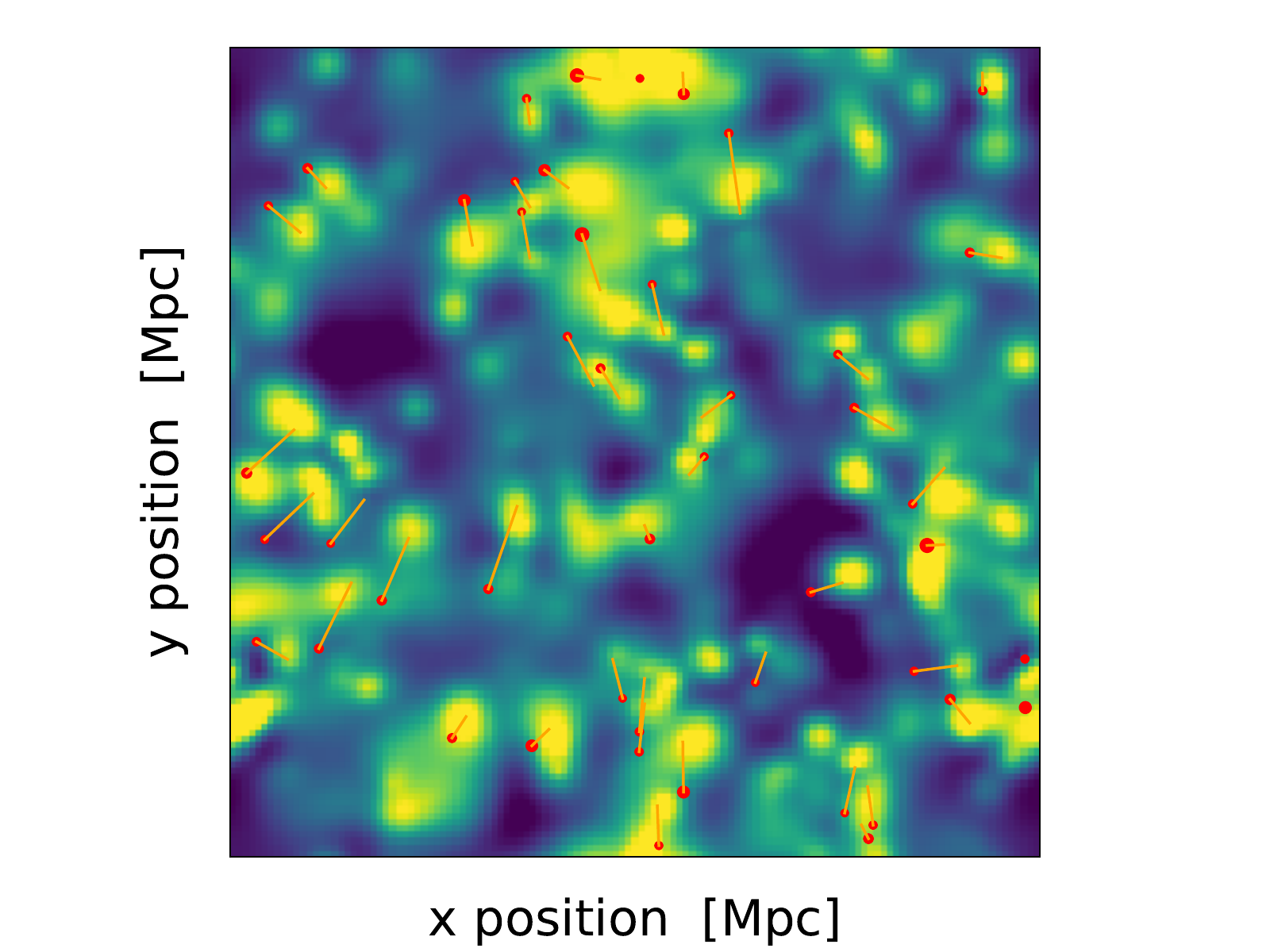}
\end{center}
\caption{The results of reconstruction algorithms applied to $M>10^{12}M_\odot$ halos in a $100~$Mpc box (see Fig.~\ref{fig:algorithm} for the input field that is being reconstructed).  The left panel is our linear solution that bounds the efficacy of nonlinear reconstruction, the middle is nonlinear reconstruction in our \emph{realistic} scenario (which initializes the reconstruction with the halo field for $k<0.2$Mpc$^{-1}$), and the right panel in our \emph{idealized} scenario, for the same linear field as shown in Figure~\ref{fig:algorithm}.  Each slice is a 20 Mpc deep projection, and the image saturates at overdensities of [-3,3].  These calculations are for $127^3$ elements, matching the fiducial resolution of our $200~$Mpc box.   Halo positions, masses and displacements are illustrated in the same manner as in Fig.~\ref{fig:algorithm}. \label{fig:linearreconst}}
\end{figure}
 
\label{ss:dependences}
 \begin{figure}
\epsfig{file= 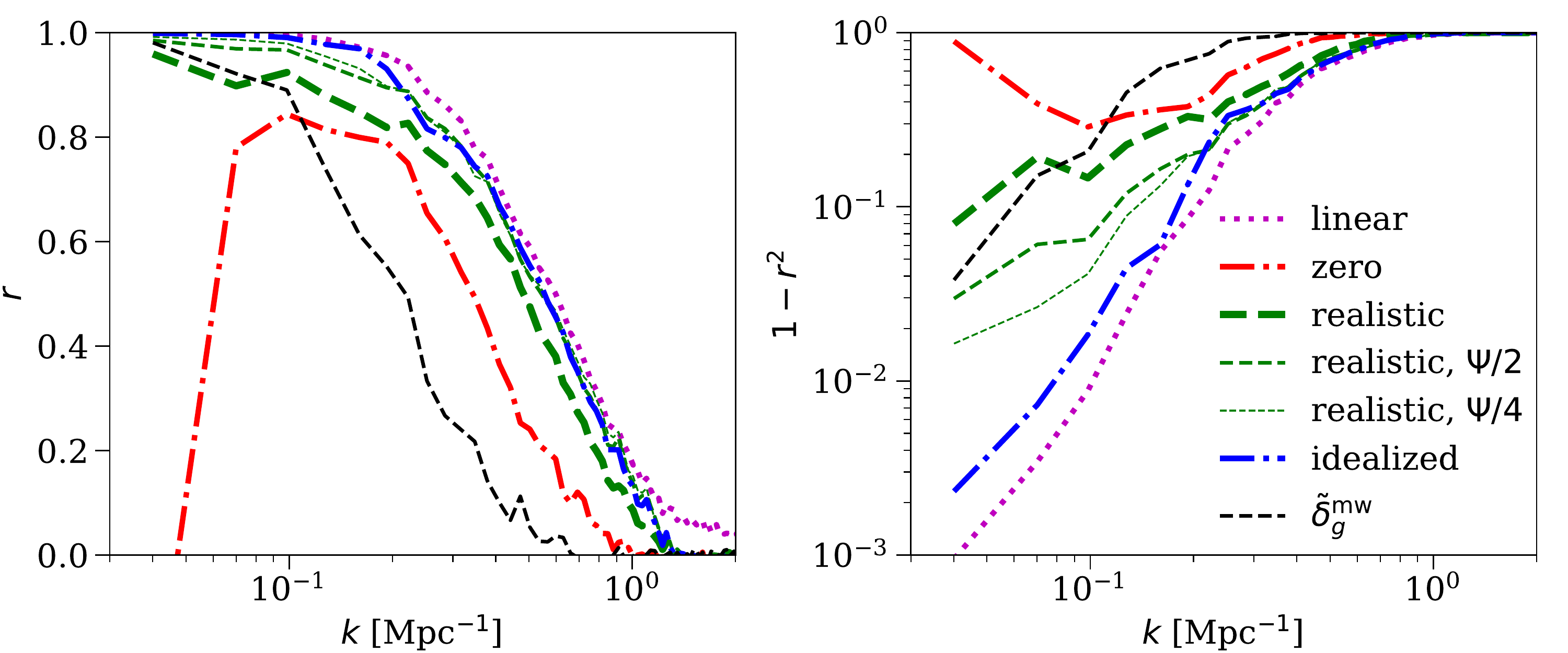, width=15cm}
\caption{{\bf Nonlinear reconstruction results:}  The cross correlation coefficients $r$ (left panel) as well as the $1-r^2$ (right panel) between the linear input overdensity and the nonlinear reconstruction for the three different initialization methods discussed in \S~\ref{sec:nonlinearmodel}.  The halo field used for these reconstructions includes all halos with $M>5\times10^{12}M_\odot$, resulting in $\bar n = 5\times10^{-4}$Mpc$^{-3}$.   The nonlinear reconstructions take $127^3$ elements in a $200\;$Mpc box, lower resolution than our linear calculations owing to their computational cost.   Also shown for comparison is the linear reconstruction approximation (using the Lagrangian overdensity and displacement constraints with $\sigma=10^{-2}$), and our \emph{realistic} algorithm but with the displacements reduced by two or four to make the problem more convex.  Finally, the black dashed curve shows the cross correlation coefficient of the mass-weighted galaxy survey with the input field: improved performance relative to this curve indicates an algorithm is outperforming the traditional shot noise floor.  
\label{fig:nonlinear}
}
\end{figure}

 Figure~\ref{fig:linearreconst} shows images of the reconstruction for the \emph{realistic} and \emph{idealized} algorithms (middle and right panels respectively), alongside the reconstruction for the linear algorithm considered in the previous section (left panel).  
    The reconstructed field appears similar between the linear algorithm and the \emph{idealized} one.  The \emph{realistic} algorithm captures many of the gross features of the two other reconstructions.  However, there are noticeable failures in the \emph{realistic}  algorithm where halos' true point of origination (as indicated by the orange line) is not close to any density peak, suggesting that the displacement is being significantly misestimated.  Furthermore, the \emph{realistic} algorithm's large-scale underdense regions are not as well captured as in the linear and idealized algorithms.  
 
 These similarities and differences between the reconstructions are also reflected in the cross correlation coefficients with the input linear overdensity field, which are shown in the left panel in Fig.~\ref{fig:nonlinear}.  The \emph{zero} algorithm does not achieve cross correlation coefficients much greater than $0.8$, and it even fails to reconstruct the density modes at all in the lowest wavenumber bin.  (Still, the \emph{zero} algorithm is able to move a long way from its trivial starting point.) The \emph{realistic} algorithm reconstructs to nearly the same maximum wavenumber as the linear one, but undershoots the precision of the linear algorithm at lower wavenumbers.  A similar undershoot relative to our linear solutions was seen in the $r$ reported in others' nonlinear reconstructions \cite{yu17,modi18} and is attributed to shot noise.  We also show the \emph{realistic} algorithm, where we have reduced the displacements by a factor of two and four, which makes the problem more convex.  The reconstruction for these cases is somewhat improved.  The \emph{idealized} algorithm, which unrealistically starts with the input field at $k<0.2\;$Mpc$^{-1}$ but where the starting point at higher wavenumbers is far from the truth, nearly saturates the bounds set by our linear calculations.  

One of the central unanswered questions is whether the shot noise of the galaxy field limits the low-wavenumber performance of these algorithms or whether the limit achieved by the linear algorithm is achievable.  The righthand panel in Fig.~\ref{fig:nonlinear} features the low wavenumber behavior, showing $1-r^2$.  Our \emph{idealized} algorithm is able to nearly match the performance of the linear algorithm.  
  It is interesting that \emph{idealized}  reconstruction is not pulled significantly away from the very small errors that are achieved by the linear algorithm at low wavenumbers, especially since our starting point at $k\gtrsim 0.2$Mpc$^{-1}$ is still fairly distant from the true solution.  However, our \emph{realistic} algorithm's solution is not able to achieve as small errors at low $k$, having similar noise to the mass-weighted halo field that this algorithm's lowest $k$ modes were initialized with.  Compare with the black dashed curve, which shows $1-r^2$ between the input linear overdensity and the mass-weighted halo field.   If we artificially reduce the displacements by a factor of two or four, making the equations we are solving more convex, $r$ does decrease significantly below the $r$ of the mass-weighted halo field that the algorithm was initialized with.  (This is despite the level of shot noise being \emph{larger} in these reduced displacement cases since the field is less clustered.)  The standard conception of shot noise has little to do with the magnitude of displacements and so the reduced noise when we reduce the displacement normalization to us suggests shot noise is not a fundamental limit (and this exercise may have some physical relevance, as at high redshifts the displacements are indeed smaller).\footnote{We also find that if the less constrained singular vectors are dampened by choosing $\sigma \sim 1$, the improvement in $r$ at low wavenumbers is similar to decreasing the displacements by a factor of $2$.}  We suspect that algorithmic improvements for solving the nonlinear equations may allow one to outperform shot noise in a realistic setting, as there are clear deficiencies in the reconstructed field of the \emph{realistic} algorithm.    


\section{Conclusions}

We have considered galaxy reconstruction in a simplified model for structure formation in which halos are the displaced Lagrangian peaks in the density field, a model that results in a deterministic nonlinear relation between the input and evolved fields.  This controlled setting allowed us to investigate the ultimate limits of reconstruction, the effects of the standard Gaussian prior on mode amplitudes, and why gradient descent-like reconstruction algorithms work at all (as it was not obvious to us why they do not get stuck far from the true solution).  For much of this study, we considered an intuitive linearized limit in which reconstruction is a convex problem but where the answer is also a solution to our nonlinear problem -- a limit that bounds the effectiveness of reconstruction. Key findings include:
\begin{itemize}
\item Existing nonlinear reconstruction algorithms are close to extracting all of the accessible information.  This argument rest on the tenet that our model of displaced Lagrangian peaks captures the essential information.  We can then `linearize' this model to make the problem convex (resulting in linear equations specifying that the modes at halo positions sum to an overdensity of $1.7$ and to the Zeldovich displacements) and find the global solutions that would clearly be the best solution an algorithm applied to our full nonlinear model could hope to find.  Despite the linearized model using information beyond what any observer can access, specifically the true displacements at the positions of the galaxies, we showed that the linearized solutions produce similar cross correlation coefficients to those of nonlinear reconstruction algorithms applied to cosmological N-body simulations \cite[][particularly when we restrict to just the constraints from displacements]{yu17,modi18} as well as to the best solutions we obtain for our full nonlinear problem. 
\item Galaxy displacements generally drive the efficacy of reconstruction rather than other properties (such as the height of peaks that collapse to form halos). This result supports why the displacement reconstruction of \cite{yu17} appears to be as successful as other algorithms.   Additional properties beyond displacements -- which are weighted to higher wavenumbers where the problem is more under-constrained -- contribute an $r<0.5$ tail to high wavenumbers in many of the mock surveys we investigated.  Extracting cosmological constraints from this tail may be challenging.  
\item The displacements constrain $\approx 3N$ independent numbers at a cosmologically interesting level for a regularization that mimics the Gaussian prior used by reconstruction algorithms, where $N$ is the number of galaxies (and an analogous conclusion applies to the $N$ Lagrangian overdensity conditions).  
   While it might seem that our setup of $\approx 3N$ equations drives this result, it did not have to be the case that the problem was sufficiently well conditioned to result in $3N$ cosmologically meaningful constraints.  We showed that the conditioning was such that most of these $3N$ numbers could be reconstructed with reasonable errors on the displacements, errors that could be achieved with our \emph{realistic} nonlinear reconstruction algorithm.  
  The effectiveness of how a particular mode is reconstructed depends on whether it projects onto the set of well-constrained eigenvectors.   For displacements, these well-constrained eigenvectors are roughly approximated by the $3N$ lowest wavenumber Fourier modes and, thus, the scale that sets the rather abrupt transition from where the cross correlation coefficient  goes from one to zero is well approximated by where the number of constraints is equal to the number of modes.  This constraint-counting wavenumber we showed is strikingly similar to the wavenumber where shot noise begins to dominate the power in the concordance cosmology across both redshift and halo mass threshold, possibly explaining why others had attributed this transition to shot noise.  
\item  We found that if galaxy displacements can be sufficiently well constrained, the $1-r^2$ of the input field with the reconstructed field falls below the shot noise expectation at low wavenumbers.  For the standard Gaussian prior on mode amplitudes, the limiting standard deviation is a factor of $\sim (k/[1~{\rm Mpc}^{-1}])^2$ smaller than the naive shot noise expectation.  This result was derived with a model where the displacements are Zeldovich and phrased in terms of the error on the displacement, a displacement which a reconstruction algorithm would have to model.  We could only reproduce this behavior when solving our full nonlinear problem when starting in an unphysical manner where we initialized the least squares solver with the input overdensity field at low wavenumbers.  If we instead initiated the reconstruction in a more realistic way that uses the galaxy field as our starting point, a precision similar to the standard shot noise limit resulted.   However, visually this reconstructed field has obvious failures, indicating to us that improvements may be possible.  
 \item Our nonlinear setup provides intuition into why reconstruction is successful:  The displacements from scales where the galaxy field does not correlate well with the linear overdensity field tend to be comparable or smaller than the Lagrangian halo size -- with ``smaller than'' being the direction where reconstruction becomes a convex problem.  As a result of this convexity, our nonlinear reconstruction was almost able to saturate the bounds set from our linear reconstruction from a starting point that is far from the true nonlinear field.  
\end{itemize}
We hope to analyze our toy nonlinear problem in future work, especially to better understand the sensitivity limit at low wavenumbers.  Understanding the effect of redshift space distortions and of modeling imperfections owing to astrophysics in our simplified setup are directions that may also merit study.


The jury is still out as to whether fully nonlinear reconstruction algorithms can be applied in a controlled manner to large-scale structure data.  As a point of optimism, we note that the situation was similar for the perturbation theory of large-scale structure until recently, when theories were developed to the point where they could be applied to SDSS/BOSS galaxy power spectrum measurements \citep{2019arXiv190905277I, 2019arXiv190905271D}.  However,  perturbation theory is under better control than the ungainly methods used by nonlinear reconstruction algorithms.  One wonders if there is some reduction of the problem that allows more control, an example being the simpler-to-understand displacement reconstruction of \cite{yu17}.    It would be a shame if analyses of spectroscopic galaxy surveys were limited to low order statistics on perturbative scales -- we understand the dynamics of structure formation on scales well beyond where the perturbative solutions are applicable.

Another unresolved issue regards what limits the precision of reconstruction at low wavenumbers.  The finding in much the literature is that it is something shot noise-like.  
  Shot noise sets the limit when one Fourier transforms a gridded halo field, but even there it is mitigated to the extent halos trace the comic mass distribution \citep{2009PhRvL.103i1303S, 2010PhRvD..82d3515H, 2018arXiv181110640S}.  However, in the setup of the problem discussed here (where there is a deterministic model that relates the input to a final field), it is far from apparent how shot noise should manifest in the reconstructed field.  In the limit where one can sufficiently constrain the displacement of each galaxy, we showed that the error could be orders of magnitude smaller than the traditional shot estimate at low wavenumbers and in a manner that is not white noise-like.  
  Yet, we did not resolve whether galaxy reconstruction could achieve these errors in a realistic setting.   
  A significant reduction over shot noise would enable better constraints on primordial non-Gaussianity and neutrino masses and so we think that pursing a more fundamental understanding is worthwhile.  

\acknowledgments
We thank Yen-Chi Chen,  Avery Meiksin, Uros Seljak, Blake Sherwin, Martin White and Jennifer Yeh for useful conversations.   We thank Vid Ir\v si\v c, Marcel Schmittfull, and especially Joanne Cohn for useful comments on an earlier version of this manuscript. We acknowledge support from NSF AAG award  2007012. 

\bibliographystyle{JHEP}
\bibliography{References}
\end{document}